\documentclass[printer]{aa}
\usepackage{graphicx}

\usepackage{natbib}
\bibpunct{(}{)}{;}{a}{}{,} 

\usepackage{txfonts}

\newcommand{\Teff}{\textrm{T}_{\textrm{eff}}} 
\newcommand{\logg}{\log~g}

\begin{document}

\title{
Abundances and kinematics for ten anticentre open clusters
}

   \subtitle{}

\author{
T. Cantat-Gaudin\inst{\ref{UNIPD},\ref{OAPD}}
\and
P. Donati \inst{\ref{OABO},\ref{UNIBO}}
\and
A. Vallenari\inst{\ref{OAPD}}
\and
R. Sordo\inst{\ref{OAPD}}
\and
A. Bragaglia\inst{\ref{OABO}}
\and
L. Magrini\inst{\ref{ARCETRI}}
}

\institute{
Dipartimento di Fisica e Astronomia, Universit\`a di Padova, vicolo Osservatorio 3, 35122 Padova, Italy\label{UNIPD}
\and
INAF-Osservatorio Astronomico di Padova, vicolo Osservatorio 5, 35122 Padova, Italy\label{OAPD}
\and
INAF-Osservatorio Astronomico di Bologna, via Ranzani 1, 40127 Bologna, Italy\label{OABO}
\and
Dipartimento di Fisica e Astronomia, Universit\`a di Bologna, via Ranzani 1, 40127 Bologna, Italy\label{UNIBO}
\and
INAF-Osservatorio Astrofisico di Arcetri, Largo E. Fermi, 5, 50125 Firenze, Italy\label{ARCETRI}
}

\date{Received date / Accepted date }

\abstract
{Open clusters are distributed all across the Galactic disk and are convenient tracers of its properties. In particular, outer disk clusters bear a key role in the investigation of the chemical evolution of the Galactic disk.}
{The goal of this study is to derive homogeneous elemental abundances for a sample of ten outer disk OCs, and investigate possible links with disk structures such as the Galactic Anticentre Stellar Structure.}
{We analysed high-resolution spectra of red giants, obtained from the HIRES@Keck and UVES@VLT archives. We derived elemental abundances and stellar atmosphere parameters by means of the classical equivalent width method. We also performed orbit integrations using proper motions.}
{The Fe abundances we derive trace a shallow negative radial metallicity gradient of slope -0.027$\pm$0.007\,dex\,kpc$^{-1}$ in the outer 12\,kpc of the disk. The [$\alpha$/Fe] gradient appears flat, with a slope of 0.006$\pm$0.007\,dex\,kpc$^{-1}$. The two outermost clusters (Be~29 and Sau~1) appear to follow elliptical orbits. The cluster Be~20 also exhibits a peculiar orbit with a large excursion above the plane.}
{The irregular orbits of the three most metal poor clusters (two of which are located at the edge of the Galactic disk), are compatible with an inside-out formation scenario for the Milky Way in which extragalactic material is accreted onto the outer disk. This is the case if the irregular orbits of these clusters are confirmed by more robust astrometric measurements such as those of the Gaia mission. We cannot determine whether Be~20, Be~29, and Sau~1 are of extragalactic origin, as they may be old, genuine Galactic clusters whose orbits were perturbed by accretion events or minor mergers in the past 5\,Gyr, or they may be representants of the thick disk population. The nature of these objects is intriguing and deserves further investigation in the near future. }


\maketitle{}

\section{Introduction}
This study aims at determining chemical abundances of various elements in stars belonging to open clusters situated in the outer disk of our Galaxy.

The first use of OCs as tracers of the chemical properties of the Galactic disk was that of \citet{Janes79}, who showed the presence of a negative metallicity gradient\footnote{Throughout this paper, metallicity is synonymous with iron abundance [Fe/H].}. Janes' observation has been confirmed by many studies. A non-exhaustive list of such works include \citet{Friel95}, \citet{Twarog97}, \citet{Friel02}, \citet{Yong05}, \citet{Bragaglia06}, \citet{Magrini09}, \citet{Lepine11}, \citet{Yong12}, and \citet{Frinchaboy13}. In particular, \citet{Twarog97} noticed a bi-modality in the metallicity gradient, and suggested that the metallicity distribution could be described as a step function with a transition radius that is a few kiloparsecs beyond the solar circle.
Modern instruments have made it possible to collect high-resolution spectroscopy for large
samples of OCs, and numerous studies have confirmed the idea
that the metallicity gradient presents two separate regimes. 

The current picture suggests the inner gradient is steep \citep[-0.09 to -0.20\,dex\,kpc$^{-1}$;][]{Yong12,Frinchaboy13}, while the outer gradient is either shallow or flat, with a break
occurring somewhere between 10 and 12 kpc \citep{Carraro04,Lepine05,Villanova05,Yong05,Carraro07,Sestito08,Pancino10,Yong12,Frinchaboy13}. Numerous studies have been devoted to understanding how the evolution of the Milky Way shaped this chemical gradient \citep[e.g.][]{Chiappini01,Cescutti07,Magrini09,Lepine11}. Various physical processes are responsible for the observed distribution of elements in the Milky Way disk. Chemical evolution models explore the influence of star formation rates, recycling of material processed by the stars, infall rates of extragalactic material onto the disk, and the variation of those quantities with Galactocentric radius to determine the combination that best reproduces the shape of the gradient at various epochs. Recent works based on chemodynamical simulations \citep{Roskar08,Minchev13,Minchev14iau} take into account stellar migration, and the fact that the present-day Galactocentric radius of a tracer may be different from its birth radius.

The uncertainty surrounding the current knowledge of the chemical distribution in the Milky Way originates partly from the multitude of methods adopted by the many different studies tackling the issue of the metallicity gradient \citep[see e.g.][]{Heiter14}.
One viable way to overcome this problem resides in pursuing homogeneity; for example, it is possible to adopt the same method of analysis which can guarantee a much more robust interpretation of the observations, and is at least freed from the systematics hidden when comparing results obtained with different approaches. In this context, the advent of large spectroscopic surveys, like APOGEE \citep{Majewski15}, GALAH \citep{DeSilva15} or the Gaia-ESO Survey \citep[][]{Gilmore12,Randich13} is playing a major role.
In particular, the ongoing Gaia-ESO Survey (GES), conducted with the UVES and GIRAFFE instruments and VLT, aims at collecting and analysing high-resolution spectra for $10^5$ stars in our Galaxy, including $10^4$ targets in 70 clusters, providing a large and homogeneous dataset and facilitating new insights into the chemical gradient in the Galactic disk.

Until now, GES data releases have focused mainly on inner disk clusters \citep{Donati14,Friel14,CantatGaudin14m11} and the chemical evolution of the disk traced by those clusters \citep[][Jacobson et al., submitted]{Magrini15}. The external regions of the Galaxy will be the focus of further releases.
This present study aims at complementing the GES dataset by analysing the spectra of red giant stars in OCs that are not planned to be observed by GES and are all located in the outer disk. Those additional spectra come from the UVES@VLT and HIRES@Keck archives, and present a similar resolution and wavelength coverage to those acquired by GES. We make use of the GES spectral line list \citep{Heiter15} and model atmospheres \citep[MARCS;][]{Gustafsson08}. The analysis is conducted with the tools DOOp \citep{CantatGaudin14doop} and FAMA \citep{Magrini13fama}, are used in the data analysis of GES and deliver similar results as the final GES adopted values, especially in the analysis of red giants \citep{Smiljanic14}.

\section{A sample of ten open clusters} \label{sec:introclusters}
\begin{figure*}[ht!]
\begin{center} \includegraphics[scale=0.45]{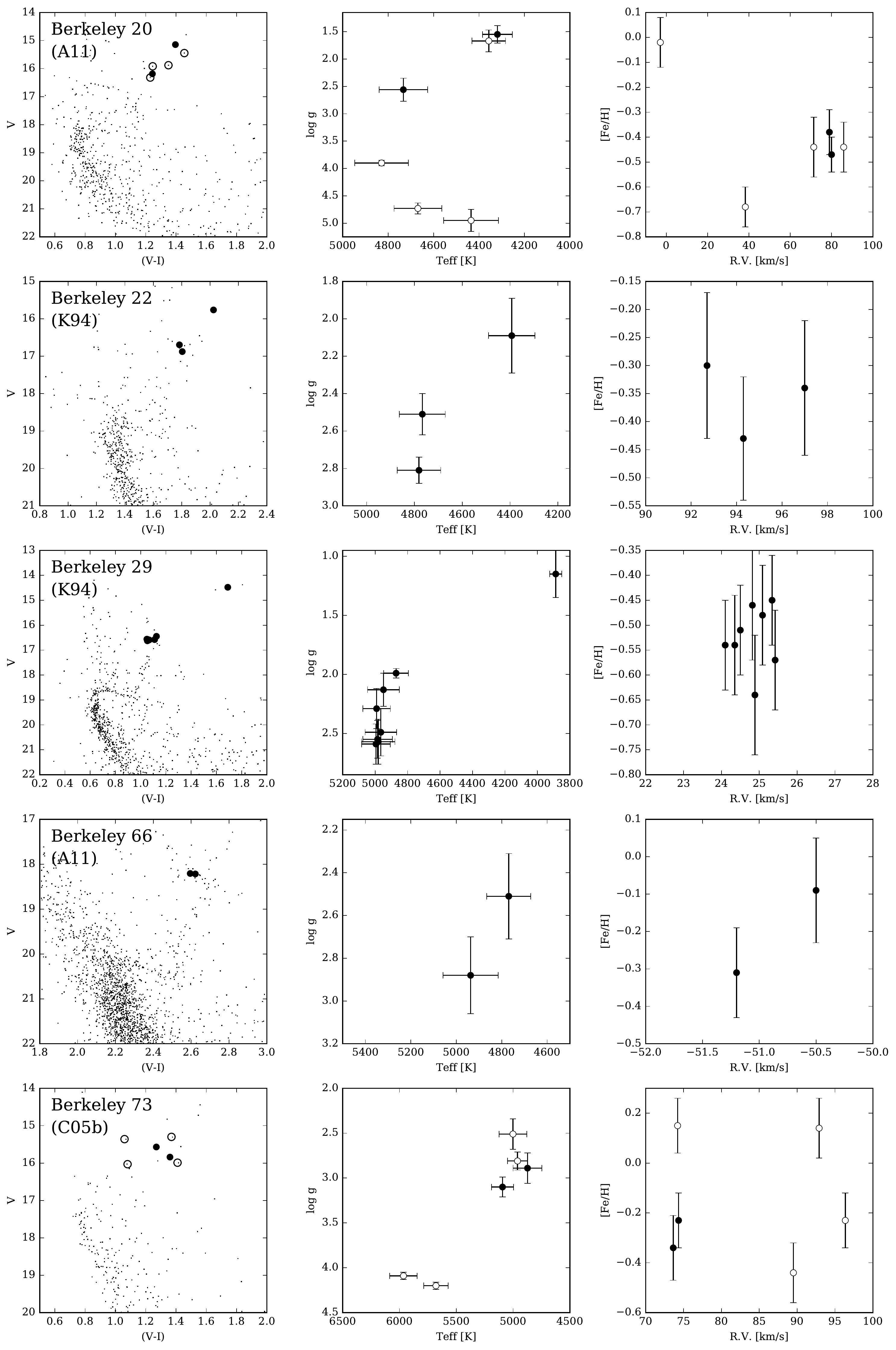} \caption{\label{fig:summary_part1} \textit{Left panels:} $VI$ CMD of Be~20, Be~22, Be~29, Be~66, and Be~73. The spectroscopic targets analysed in this study are indicated with filled dots (for cluster members) and empty dots (non-members). \textit{Middle panels:} effective temperature and surface gravity of the spectroscopic targets (as derived in this study). \textit{Right panels:} radial velocities and metallicities (both as computed in this study) for the spectroscopic targets. References for photometry: A11 \citep{Andreuzzi11}, K94 \citep{Kaluzny94}, and C05b \citep{Carraro05be}.  } \end{center}
\end{figure*}

\begin{figure*}[ht!]
\begin{center} \includegraphics[scale=0.45]{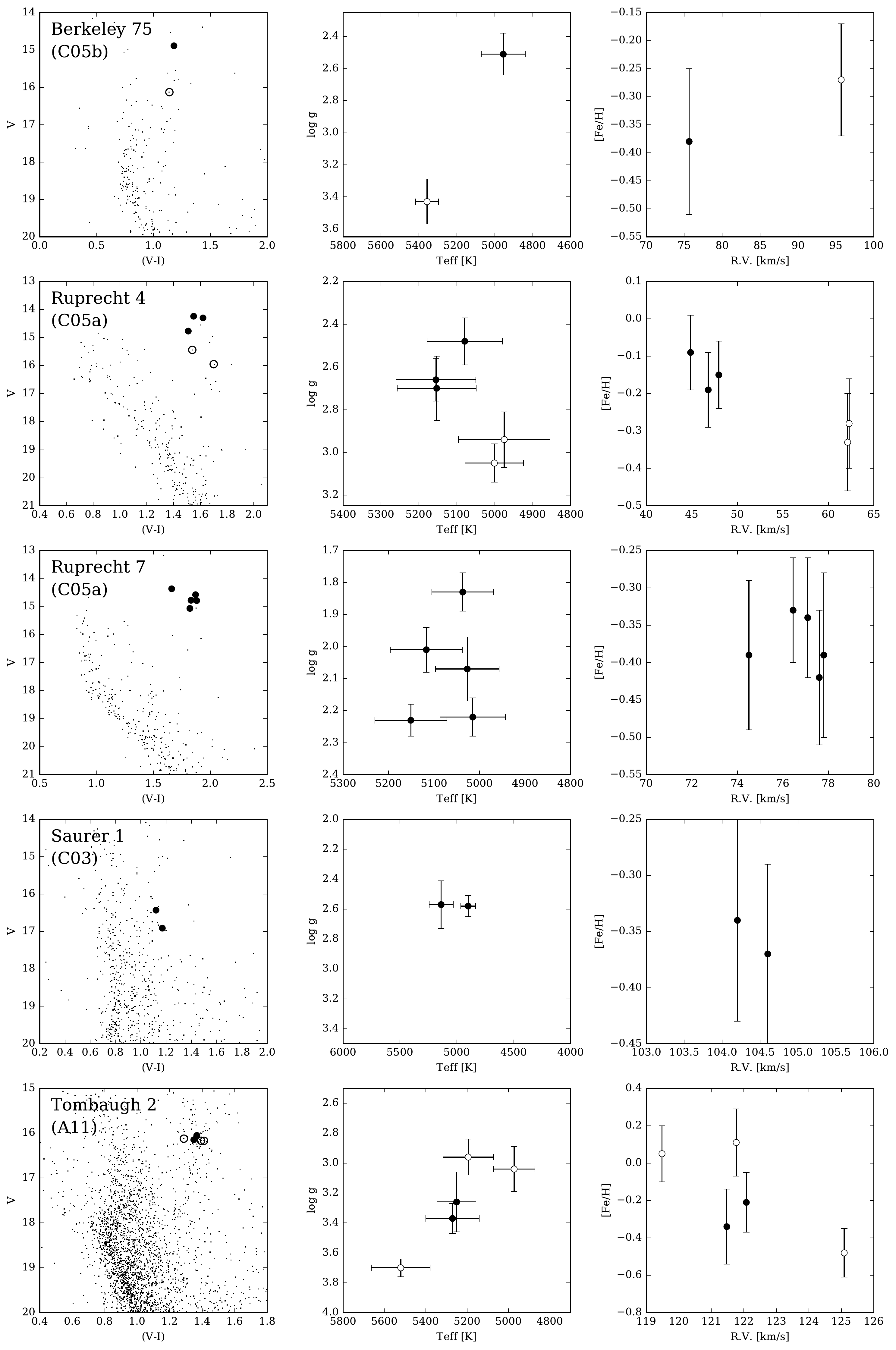} \caption{\label{fig:summary_part2} Same as Fig.~\ref{fig:summary_part1} for Be~75, Rup~4, Rup~7, Sau~1, and Tom~2. References for photometry: C05b \citep{Carraro05be}, C05a \citep{Carraro05rup}, C03 \citep{Carraro03}, and A11 \citep{Andreuzzi11}.} \end{center}
\end{figure*}

\begin{figure*}[ht]
\begin{center} \includegraphics[scale=0.85]{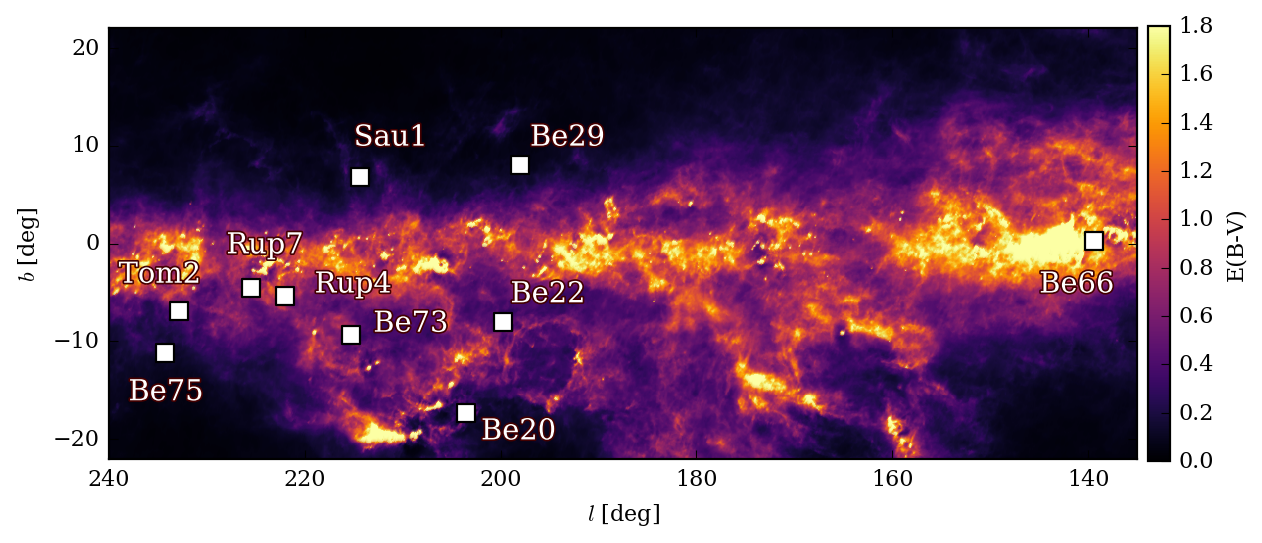} \caption{\label{fig:map_extinction_anticentre} Location (in Galactic coordinates) of the ten OCs studied in this paper. The background is the extinction map of \citet{Schlegel98}. } \end{center}
\end{figure*}

We built a sample of ten outer disk OCs for which we found publicly available high-resolution spectra. The photometry of the clusters and the spectroscopic targets is shown in Figs.~\ref{fig:summary_part1} and \ref{fig:summary_part2}. The sky distribution (in Galactic coordinates) of these clusters is shown in Fig.~\ref{fig:map_extinction_anticentre}. Their main parameters are summarised in Table~\ref{tab:outerclusters}.

The complete list of the the spectroscopic targets analysed in this paper is presented in Table~\ref{tab:targets}.


\subsection*{Berkeley 20}
The first photometric study of Be~20 is that of \citet{MacMinn94}. From $VI$ observations these authors derived an age of about 6\,Gyr, a metallicity [Fe/H]=-0.23, an extinction $E(V-I)=0.16$, and a distance modulus $(m-M)_0$=15.0 (R$_{\mathrm{GC}}$=15.8\,kpc); these findings made Be~20 at the time the most distant known OC, which also presented an unusual location of 2.5\,kpc below the Galactic disk. \citet{Durgapal01}, based on $BVRI$ photometry, found an age of 5\,Gyr, a metallicity [Fe/H]=-0.3, and a distance modulus $(m-M)_0$=15.1. More recently, \citet{Andreuzzi11} derived an age of 5.8\,Gyr, a distance modulus $(m-M)_0$=14.7 (d$_{\odot}$=8.7\,kpc, R$_{\mathrm{GC}}$=16.0\,kpc) and a metallicity Z=0.008 ([Fe/H]=-0.3).

\citet{Friel02}, based on low-resolution spectra of nine stars, identified six cluster members with radial velocities of +70$\pm$13\,km\,s$^{-1}$ and metallicity [Fe/H]=-0.61$\pm$0.14. The study of \citet{Frinchaboy06} identified five members out of 20 targets with radial velocities of +75$\pm$2.4\,km\,s$^{-1}$.

High-resolution spectroscopy was carried out by \citet[][hereafter Y05]{Yong05} who derived a mean radial velocity of +78.9$\pm$0.7\,km\,s$^{-1}$ for four stars, and a mean metallicity [Fe/H]=-0.49$\pm$0.06 for two of them. \citet[][hereafter S08]{Sestito08} obtained UVES spectra for six stars, two of which are confirmed members. They found a metallicity of \mbox{[Fe/H]=-0.3$\pm$0.02}. Our study reanalysed the data collected for those six stars by S08.

The orbit reconstructions of \citet{Wu09} show a large eccentricity for the orbit of Be~20. Combined with is low metallicity and old age, \citet{Wu09} propose Be~20 is a thick disk cluster.
On the other hand, the study of \citet{VandePutte10} attributes this peculiar orbit to an extragalactic origin.

\subsection*{Berkeley 22}
Be~22 was first studied by \citet{Kaluzny94} with $VI$ photometry, who derived an age of 3\,Gyr and a distance of 6\,kpc (R$_{\mathrm{GC}}$=13.7 with a significant extinction $E(V-I)=0.74$. The study suggests a subsolar metallicity of Z=0.008 for this cluster. \citet{DiFabrizio05} consider that the cluster may be slightly younger (2.5\,Gyr) and of solar metallicity. A subsequent study by \citet[][hereafter V05]{Villanova05} weighs in favour of the results of \citet{Kaluzny94}. The study of V05 also makes use of high-resolution spectra of two stars acquired with the HIRES spectrometer at the Keck telescope, from which they derive a cluster metallicity [Fe/H]=-0.32$\pm$0.03. \citet{Bragaglia06}, by means of $BVI$ photometry, found the age of Be~22 to be between 2.1 and 2.5\,Gyr depending on the adopted stellar models. Finally, \citet{Yong12} derive a metallicity [Fe/H]=-0.45 based on high-resolution spectroscopy of two red giants.

The present study analyses the spectra of the two V05 stars, plus the high-resolution spectrum of a third star also acquired with HIRES.

\subsection*{Berkeley 29}
Be~29 is the most distant known open cluster from the Galactic centre. The first photometric study of this object is that of \citet{Kaluzny94}, who estimated a low metallicity of [Fe/H]=-1 and a Galactocentric distance R$_{\mathrm{GC}}$=18.7\,kpc. \citet{Tosi04} used $BVI$ photometry and estimate an age between 3.4 and 3.7, a Galactocentric distance R$_{\mathrm{GC}}$=21 to 22\,kpc and a subsolar metallicity (Z=0.004 -- 0.006), while \citet{Bragaglia06} derive an age of 3.7\,Gyr.

\citet[][hereafter C04]{Carraro04} analysed HIRES high-resolution spectra of two red clump stars in this cluster and derived a metallicity [Fe/H]=-0.44$\pm$0.18, for a Galactocentric distance of R$_{\mathrm{GC}}$=21.6\,kpc and an age of 4.5\,Gyr. \citet{Carraro04} also suggest that Be~29 features enhanced $\alpha$-abundances, which is contradicted by \citet{Frinchaboy06}.
\citet{Yong05}, analysing spectra of two stars near the RGB tip, derive a metallicity [Fe/H]=-0.54 and [$\alpha$/Fe]=+0.2.  A study by S08 using UVES high-resolution spectra of 6 targets yielded a mean value [Fe/H]=-0.31$\pm$0.03 and no significant enhancement in [$\alpha$/Fe].

Orbit integrations were performed by \citet{VandePutte10} who find that Be~29 is likely to be counter-rotating in the plane of the Galaxy. They also confirm the result of \citet{Wu09} who determine that the orbit of Be~29 extends to high altitudes above the Galactic plane.

The present study analyses the HIRES high-resolution spectra of the two stars studied by C04, as well as six targets observed with the UVES spectrometer and studied by S08. One of thse targets happens to be in common between both samples (star K801 in C04 is the same as star T398 of S08).

\subsection*{Berkeley 66}
\citet{Phelps96} made use of $VI$ photometry to derive an age of
3.5\,Gyr, a total distance modulus $(m-M)_0$=17.4 with a significant extinction $E(V-I)=1.6$ (d$_{\odot}$=5.2\,kpc, R$_{\mathrm{GC}}$=12.9\,kpc), and a subsolar metallicity in the range [Fe/H]=-0.2 -- 0\,dex.
 \citet{Andreuzzi11} acquired new $BVI$ photometry, from which they determined that the metallicity must be subsolar. They derived a distance modulus $(m-M)_0$=13.3 (d$_{\odot}$=4.6\,kpc, R$_{\mathrm{GC}}$=11.9\,kpc), and an age of 3.8\,Gyr.

\citet{Villanova05} obtained HIRES spectra of two red clump stars and derived a metallicity value for one of these stars of \mbox{[Fe/H]=-0.48$\pm$0.24\,dex}. Our study reanalyses the spectra of both targets of V05.

\subsection*{Berkeley 73}
The first photometric study of this cluster is that of \citet{Ortolani05}, who derived an age of 2.3\,Gyr from $BV$ photometry, a metallicity Z=0.008, and distance modulus $(m-M)_0$=14.7 (d$_{\odot}$=6.5\,kpc, R$_{\mathrm{GC}}$=10.2\,kpc).
From $BVI$ photometry, \citet[][hereafter C05b]{Carraro05be} derived a significantly different age of 1.5\,Gyr, a metallicity Z=0.008, and therefore a larger distance modulus $(m-M)_0$=15.3 (d$_{\odot}$=9.7\,kpc, R$_{\mathrm{GC}}$=16.4\,kpc).

\citet[][hereafter C07]{Carraro07} analysed UVES spectra for six stars in Be~73. Considering two of these stars as probable cluster members yields a mean metallicity [Fe/H]=-0.22$\pm$0.10. We reanalysed the C07 data for those six stars.

\subsection*{Berkeley 75}
The $BVI$ photometric study of C05b provided an age of 3\,Gyr, Z=0.004, and $(m-M)_0$=15.2 (d$_{\odot}$=9.8\,kpc, R$_{\mathrm{GC}}$=16.2\,kpc).

\citet{Carraro07} observed two Be~75 stars with UVES and concluded that only one of these stars is a possible cluster member with a metallicity [Fe/H]=-0.22$\pm$0.20 (Z=0.011). A higher metallicity would mean this cluster is in fact significantly closer than previously determined by C05b with a revised distance modulus $(m-M)_0$=14.9 (d$_{\odot}$=9.1\,kpc, R$_{\mathrm{GC}}$=15.5\,kpc). The present study reanalyses the available UVES spectra for those two stars.

\subsection*{Ruprecht 4}
The $BVI$ photometry study of \citet[][hereafter C05a]{Carraro05rup} found an age of 0.8\,Gyr, Z=0.008, and $(m-M)_0$=14.3 (d$_{\odot}$=4.9\,kpc, R$_{\mathrm{GC}}$=12.0\,kpc).

\citet{Carraro07} analysed UVES spectra for five stars in this cluster, three of which can be considered probable cluster members and yield a mean metallicity [Fe/H]=-0.09$\pm$0.05.

\subsection*{Ruprecht 7}
The first study of this object was presented by \citet{Mazur93}, who determined an age of 0.7\,Gyr, a distance modulus $(m-M)_0$=15.5 and a low metallicity Z=0.004.
The $BVI$ photometry study of C05a obtained an age of 0.8\,Gyr and $(m-M)_0$=15.0 (d$_{\odot}$=6.5\,kpc, R$_{\mathrm{GC}}$=13.8\,kpc). \citet{Carraro05rup} also determined a significantly different metallicity, finding that isochrones of solar metallicity provide a better fit to the cluster CMD.

\citet{Carraro07} analysed UVES spectra for five stars in this cluster, which are all considered cluster members and have a mean metallicity [Fe/H]=-0.26$\pm$0.05. We reanalyse the spectra of these five targets.

\subsection*{Saurer 1}
This cluster is one of the most distant known OCs, as well as one of the oldest. \citet{Frinchaboy02} estimate an age of 7$\pm$3\,Gyr and Galactocentric distance R$_{\mathrm{GC}}$=19$\pm$1\,kpc, from $VI$ photometry. 
Using $VI$ photometry as well, \citet{Carraro03} determined an age of 5\,Gyr and a Galactocentric distance R$_{\mathrm{GC}}$=19.2\,kpc. 

A subsequent spectroscopic study by C04 targeting two red clump stars derived a mean metallicity of [Fe/H]=-0.38$\pm$0.14. Those two targets are reanalysed in this study.

\subsection*{Tombaugh 2}

The oldest data available for this clusters is the $UBV$ CMD of \citet{Adler82}.
The $BVI$ photometric study by \citet{Kubiak92} estimated an age of 4\,Gyr and a distance d$_{\odot}$=6.3$\pm$0.9\,kpc.
Later, \citet{Phelps94} determined an age of 2.5Gyr.

\citet{Brown96} analysed medium-resolution spectra of five stars and derived a metallicity [Fe/H]=-0.4$\pm$0.25.
\citet{Friel02} obtained low-resolution spectra for 12 stars and derived a metallicity [Fe/H]=-0.44$\pm$0.09.

\citet{Frinchaboy08} present high-resolution spectroscopy of five red clump stars observed with UVES and derive an age of 2\,Gyr and d$_{\odot}$=7.9\,kpc using the photometry of \citet{Phelps94} and radial velocity members. \citet{Frinchaboy08} derive a metallicity [Fe/H]=-0.07$\pm$0.01.
They also acquired medium-resolution GIRAFFE spectra, which revealed a group of metal poor stars with metallicities down to [Fe/H]=-0.28, and suggest Tom~2 could feature an internal metallicity spread. Another explanation may be that the metal poor sample belong to a surrounding stream associated with Tom~2, and therefore have similar radial velocities but a different chemistry.

\citet{Villanova10} studied the GIRAFFE spectra of 37 targets and derived an average [Fe/H]=-0.31$\pm$0.01 for 13 members, with no visible abundance spread.

A more recent $BVI$ photometric study by \citet{Andreuzzi11} obtained an age of 1.6\,Gyr, half solar metallicity, and $(m-M)_0$=14.5 (R$_{\mathrm{GC}}$=14.2\,kpc).

The present study reanalyses the high-resolution spectra of the five targets of \citet{Frinchaboy08}.

\begin{table*}
\begin{center}
	\caption{ \label{tab:outerclusters} Summary of main parameters for the ten clusters under study.}
	\small\addtolength{\tabcolsep}{-3pt}
	\begin{tabular}{c c c c c c c c c c c c c c c c c}
	\hline
	\hline
	OC & $\alpha$ & $\delta$ & RA & DEC & $l$ & $b$ & d$_{\odot}$ & R$_{\mathrm{GC}}$ & z & [Fe/H] & $\sigma$[Fe/H] & [$\alpha$/Fe] & $\sigma$[$\alpha$/Fe] & members & age & age\\
	   & (J2000) & (J2000) & (hms) & (hms) & & & [kpc]     & [kpc]             & [kpc] &     & &  &  & /targets & [Gyr] & Ref. \\
	\hline
	Be~20 &   83.250 &    0.217 & 5 33 00 & 0 13 00 &  203.505 &  -17.275 & 8.7 & 16.0 & -2.5	& -0.42 & 0.08	& 0.15 & 0.13 & 2/6& 5.8 & A11\\
	Be~22 &   89.600 &    7.833 & 5 58 24 & 7 50 00 &  199.803 &   -8.052 & 6.0 & 13.7 & -0.8	& -0.36 & 0.07	& 0.11 & 0.10 & 3/3& 3.3 & V05\\
	Be~29 &  103.325 &   16.917 & 6 53 18 & 16 55 00 &  197.984 &    8.025 & 13.1 & 21.6 & 1.9	&  -0.52 & 0.03	& 0.13 & 0.12 & 7/7& 4.5 & C04\\
	Be~66 &   46.075 &   58.767 & 3 04 18 & 58 46 00 &  139.434 &    0.218 & 5.0 & 12.7 & 0.1	& -0.21 & 0.09	& 0.04 & 0.10 & 2/2& 4.7 & V05\\
	Be~73 &  95.525  &  -6.317  & 6 22 06 & -6 19 00 &  215.259 &    -9.387 & 9.7 & 16.4 & -1.6	& -0.28 & 0.08	& -0.01 & 0.11 & 2/6& 1.5 & C07\\
	Be~75 &  102.246 &  -23.992 & 6 48 59 & -23 59 30 &  234.299 &  -11.193 & 9.1 & 15.5 & -1.7	& -0.38 & 0.14	& 0.24 & 0.17 & 1/2& 4 & C07\\
	Rup~7 &  104.467 &  -13.224 & 6 57 52 & -13 13 25 &  225.449 &   -4.589 & 6.5 & 13.8 & -0.5	& -0.37 & 0.04	& 0.02 & 0.07 & 5/5& 0.8 & C07\\
	Rup~4 &  102.225 &  -10.533 & 6 48 54 & -10 32 00 &  222.047 &   -5.339 & 4.9 & 12.0 & -0.4	& -0.14 & 0.06	& 0.03 & 0.12 & 3/5& 0.8 & C07\\
	Sau~1 &  109.575 &    1.887 & 7 18 18 & 1 53 12 &  214.317 &    6.836 & 13.2 & 19.2 & 1.6	& -0.36 & 0.06	& 0.08 & 0.04 & 2/2& 5 & C04\\
	Tom~2 &  105.771 &  -20.817 & 7 03 05 & -20 49 00 &  232.832 &   -6.880 & 7.9 & 14.2 & -0.9	& -0.27 & 0.13	& 0.23 & 0.24 & 2/5& 2 & A11\\
	\hline\end{tabular}
\tablefoot{The R$_{\mathrm{GC}}$ values are computed using the literature value for d$_{\odot}$ and R$_{\mathrm{GC,\odot}}=8.3$\,kpc. The [Fe/H] and [$\alpha$/Fe] values are those derived in Sect.~\ref{sec:results} of this study. References: C04 \citep{Carraro04}, V05 \citep{Villanova05}, C07 \citep{Carraro07}, A11 \citep{Andreuzzi11}, F08 \citep{Frinchaboy08}. The locations of these clusters are shown in Fig.~\ref{fig:map_extinction_anticentre}, and Fig.~\ref{fig:gradient_Both}.
}
\end{center}
\end{table*}

\section{The data}

\subsection{Data reduction}
The spectra under analysis come from two different sources: the Keck-HIRES archive and the ESO-UVES archive. Information on the target stars we analysed is summarised in Table~\ref{tab:targets}.

\begin{table*}
\begin{center}
	\caption{ \label{tab:targets} Summary of observations for 44 stars observed in ten clusters.}
	\small\addtolength{\tabcolsep}{-1pt}
	\begin{tabular}{c c c c c c c c c c c}
	\hline
	\hline
	OC & star* & RA & DEC & $V$ & $V-I$ & date$\dagger$ & exposure time [s] & S/N & R.V. [km\,s$^{-1}$] & instrument\\
	\hline

	Be~20 &	S001201	 &  05 32 36.774 &  	 +00 11 04.84 &   	16.177 & 1.25  		& 2006-02 & 7$\times$2774 & 28	& 79.0 $\pm$ 0.6& UVES\\
	Be~20 &	S001240	 &  05 32 38.963 &  	 +00 11 20.37 &   	15.154 & 1.43  		& 2006-02 & 7$\times$2774 & 49	& 80.0 $\pm$ 0.5& UVES\\
	Be~20 &	S001403	 &  05 32 5.609 &   	 +00 12 44.04 &   	15.88 & 1.35 		& 2006-02 & 7$\times$2774 & 35	& -2.9 $\pm$ 0.6& UVES\\
	Be~20 &	S001519	 &  05 32 54.960 &  	 +00 14 07.30 &   	16.32 & 1.23  		& 2006-02 & 7$\times$2774 & 28	& 71.4 $\pm$ 1.7& UVES\\
	Be~20 &	S001666	 &  05 32 46.417 & 	 +00 15 52.19 &   	15.919 & 1.25  		& 2006-02 & 7$\times$2774 & 33	& 85.9 $\pm$ 0.7& UVES\\
	Be~20 &	S001716	 &  05 32 50.057 &  	 +00 16 16.43 &   	15.447 & 1.45  		& 2006-02 & 7$\times$2774 & 50	& 38.2 $\pm$ 0.5& UVES\\
	\hline

	Be~22 & K400 & 05 58 30.97 & +07 46 15.3 & 16.695 & 1.784 & 2004-12-01 & $3\times1800$ & 61	& 92.7 $\pm$ 0.3& HIRES\\
	Be~22 & K579 & 05 58 25.78 & +07 45 31.2 & 16.876 & 1.804 & 2004-12-01 & $3\times1500$ & 56	& 97.0 $\pm$ 0.8& HIRES\\
	Be~22 & K414 & 05 58 25.92 & +07 46 01.2 & 15.762 & 2.024 & 2006-12-28 & $2\times2400+981$ & 100	& 94.3 $\pm$ 0.4& HIRES\\  
	\hline

	Be~29 & K801  & 06 53 08.07  & +16 55 40.53 & 16.579 & 1.061 & 2004-01-14 & $3\times2400$ & 61	& 24.5 $\pm$ 0.2& HIRES\\
	Be~29 & K1032 & 06 53 03.50  & +16 55 08.50 & 16.561 & 1.049 & 2004-01-14 & $3\times2400$ & 57	& 24.1 $\pm$ 0.2& HIRES\\
	Be~29 & T159  & 06 53 01.60  & +16 56 21.11 & 16.627 & 1.053 & 2006 & $8 \times 2775$ & 24	& 25.1 $\pm$ 0.9& UVES\\
	Be~29 & T257  & 06 53 04.32  & +16 55 39.37 & 16.608 & 1.06  & 2006 & $8 \times 2775$ & 28	& 24.9 $\pm$ 0.9& UVES\\
	Be~29 & T398  & 06 53 08.07  & +16 55 40.53 & 16.591 & 1.067 & 2006 & $8 \times 2775$ & 28	& 24.4 $\pm$ 0.8& UVES\\
	Be~29 & T602  & 06 52 55.49  & +16 57 39.30 & 16.579 & 1.11  & 2006 & $8 \times 2775$ & 31	& 25.4 $\pm$ 0.8& UVES\\
	Be~29 & T933  & 06 53 04.49  & +16 57 44.69 & 16.447 & 1.125 & 2006 & $8 \times 2775$ & 31	& 25.3 $\pm$ 0.9& UVES\\
	Be~29 & T1024 & 06 53 07.13  & +16 57 12.67 & 14.458 & 1.663 & 2006 & $8 \times 2775$ & 107	& 24.8 $\pm$ 0.6& UVES\\
	\hline

	Be~66 & PJ785 & 03 04 02.90 & +58 43 57.0 & 18.21 & 2.60 & 2004-12-01 & $4\times2700$ & 47	& -50.5 $\pm$ 0.5 & HIRES\\
	Be~66 & PJ934 & 03 04 06.41 & +58 43 31.0 & 18.22 & 2.62 & 2004-12-01 & $3\times2700+1335$ & 49	& -51.2 $\pm$ 0.9 & HIRES\\ 
	\hline

	Be~73 &	CC12 &  06 21 54.89 & -06 17 51.91 & 15.30 & 1.37  			& 2005-10-31 & 2759+3059 & 35	& 89.5 $\pm$ 0.7 & UVES\\
	Be~73 &	CC13 &  06 21 55.36 & -06 20 10.04 & 15.36 & 1.06  			& 2005-10-31 & 2759+3059 & 35	& 74.2 $\pm$ 0.8 & UVES\\
	Be~73 &	CC14 &  06 22 10.44 & -06 19 30.19 & 15.57 & 1.27  			& 2005-10-31 & 2759+3059 & 23	& 74.3 $\pm$ 0.6 & UVES\\
	Be~73 &	CC15 &  06 22 01.13 & -06 19 17.14 & 15.84 & 1.36 			& 2005-10-31 & 2759+3059 & 29	& 73.6 $\pm$ 0.8 & UVES\\
	Be~73 &	CC18 &  06 22 01.99 & -06 19 21.89 & 15.99 & 1.41 			& 2005-10-31 & 2759+3059 & 25	& 96.4 $\pm$ 0.7 & UVES\\
	Be~73 &	CC19 &  06 21 57.24 & -06 17 15.55 & 16.03 & 1.08 			& 2005-10-31 & 2759+3059 & 22	& 92.9 $\pm$ 1.6 & UVES\\
	\hline

	Be~75 &	CC9   &  06 49 07.09 & -23 59 44.94 & 14.89 & 1.18  		& 2005-12-05 & 2$\times$2759 & 	50	& 75.6 $\pm$ 0.6 & UVES\\
	Be~75 &	CC22  &  06 48 55.85 & -24 00 07.16 & 16.13 & 1.14  		& 2005-12-05 & 2$\times$2759 & 	15	& 95.7 $\pm$ 1.4 & UVES\\
	\hline

	Rup~4 &	CC18	 &  06 48 51.69 & -10 30 20.49 & 14.24 & 1.55 		& 2005-12-01 & 2759 & 30	& 62.1 $\pm$ 0.7 & UVES\\
	Rup~4 &	CC29	 &  06 49 00.80 & -10 31 57.81 & 14.30 & 1.62  		& 2005-12-01 & 2759 & 26	& 62.3 $\pm$ 0.8 & UVES\\
	Rup~4 &	CC3	 &  06 48 55.37 & -10 31 32.39 & 14.77 & 1.51 		& 2005-12-01 & 2759 & 49	& 48.0 $\pm$ 0.6 & UVES\\
	Rup~4 &	CC4	 &  06 48 54.51 & -10 33 21.32 & 15.44 & 1.54 		& 2005 & 2$\times$2759 & 76	& 44.9 $\pm$ 0.6 & UVES\\
	Rup~4 &	CC8	 &  06 48 57.87 & -10 30 27.09 & 15.95 & 1.7  		& 2005-12-01 & 2759 & 44	& 46.8 $\pm$ 0.6 & UVES\\
	\hline

	Rup~7 &	CC2	 &  06 57 50.14 & -13 13 39.93 & 14.07 & 1.36  		& 2005-11 & 2$\times$2759 & 69	& 77.8 $\pm$ 1.1 & UVES\\
	Rup~7 &	CC4	 &  06 57 43.08 & -13 12 17.41 & 14.58 & 1.87  		& 2005-11 & 2$\times$2759 & 59	& 76.4 $\pm$ 0.6 & UVES\\
	Rup~7 &	CC5	 &  06 57 54.56 & -13 13 59.18 & 14.78 & 1.83  		& 2005-11 & 2$\times$2759 & 42	& 77.6 $\pm$ 1.2 & UVES\\
	Rup~7 & CC6	 &  06 57 53.66 & -13 12 57.83 & 14.79 & 1.88  		& 2005-11 & 2$\times$2759 & 61	& 77.1 $\pm$ 1.1 & UVES\\
	Rup~7 &	CC7	 &  06 57 52.85 & -13 12 55.08 & 15.07 & 2.82  		& 2005-11 & 2$\times$2759 & 54	& 74.5 $\pm$ 1.6 & UVES\\
	\hline

	Sau~1 & C91  & 07 20 54.75 & +01 47 53.09 & 16.43 & 1.11 & 2004-01-14 & $2\times2700$ & 52	& 104.6 $\pm$ 0.3 & HIRES\\ 
	Sau~1 & C122 & 07 20 57.08 & +01 48 44.97 & 16.92 & 1.17 & 2004-01-14 & $2\times3000$ & 47	& 104.2 $\pm$ 0.3 & HIRES\\ 
	\hline

	Tom~2 &	P146	&  07 03 09.688 &  -20 45 49.21 &  16.018  & 1.367 	& 2005-12-01 & 2759 & 10 & 122.1 $\pm$ 1.5 & UVES\\
	Tom~2 &	P158	&  07 03 07.759 &  -20 46 17.44 &  16.099  & 1.341 	& 2005-12-01 & 2759 & 10 & 121.5 $\pm$ 1.2 & UVES\\
	Tom~2 &	P162	&  07 03 01.566 &  -20 47 59.76 &  16.131  & 1.318	& 2005-12-01 & 2759 & 8	& 121.8 $\pm$ 2.0 & UVES\\
	Tom~2 &	P164	&  07 03 20.480 &  -20 46 47.78 &  16.160  & 1.436 	& 2005-12-01 & 2759 & 15 & 125.1 $\pm$ 1.3 & UVES\\
	Tom~2 &	P165	&  07 03 03.496 &  -20 48 48.42 &  16.161  & 1.411 	& 2005-12-01 & 2759 & 11 & 119.5 $\pm$ 2.0 & UVES\\
	\hline

\end{tabular}
\tablefoot{* C: ID from \citet{Carraro03}; K: \citet{Kaluzny94}; T: \citet{Tosi04}; PJ: \citet{Phelps96}; CC: \citet{Carraro05rup,Carraro05be}; P: \citet{Phelps94}; S: \citet{Sestito08}.
\newline $\dagger$ Only the month (or year) is given for stars observed during multiple nights.
\newline One star in Be~29 (K801=T398) was observed with both HIRES and UVES.}
\end{center}
\end{table*}

\subsubsection*{HIRES spectra}
The data for Sau~1 and Be~29 \citep[published by][]{Carraro04} and for Be~22 and Be~66 \citep[published by][]{Villanova05} was acquired with the HIRES spectrometer \citep{Vogt94} mounted on the 10-meter telescope of the W.~M. Keck Observatory (Hawaii), providing a resolution $R\sim34,000$. We obtained pre-reduced spectra through the Keck archive\footnote{\url{http://koa.ipac.caltech.edu/cgi-bin/KOA/nph-KOAlogin}} for nine stars in total. Information on those nine targets is presented in Table~\ref{tab:targets}. The HIRES archive contains data for an additional star for Be~22 (star K414) which was not analysed by V05 but which we analysed in this study. The available spectra for Be~29 and Sau~1 cover the spectral range 5700--8200\,$\AA$, while the spectra of Be~22 and Be~66 cover the range 5300--9200\,$\AA$. Unfortunately, the range 5300--6300\,$\AA$ for the star 785 of Be~66 appears to be missing from the HIRES archive.

Pre-reduced data is stored in the HIRES archive for each available exposure in the form of individual orders. The radial velocity calibration was performed using the spectra of the radial velocity standard HD26162 for Be~29 and Sau~1 \citep[24.8\,km\,s$^{-1}$][]{Wielen99} and HD82106 \citep[29.7\,km\,s$^{-1}$][]{Udry99}, acquired by C04 and V05, respectively, during their observations.
Multiple exposures covering the same spectral range for a given star were co-added to reach a higher signal-to-noise ratio. Where the individual orders cover overlapping spectral ranges, they were merged together. We did not apply any correction for telluric lines, and we did not measure lines in the spectral ranges 6865--6950 and 7590--7700$\AA$, where many saturated telluric lines are present.

\subsubsection*{UVES spectra}
We obtained the raw UVES data for Be~20, Be~73, Be~75, Rup~4, Rup~7, Tom~2, and for six stars in Be~29 from the ESO archive\footnote{\url{http://archive.eso.org/eso/eso_archive_main.html}}, at a resolution $R\sim47,000$. The available data cover the spectral range 4800--6800\,$\AA$.

\begin{figure}[ht]
\begin{center} \resizebox{\hsize}{!}{\includegraphics[scale=1]{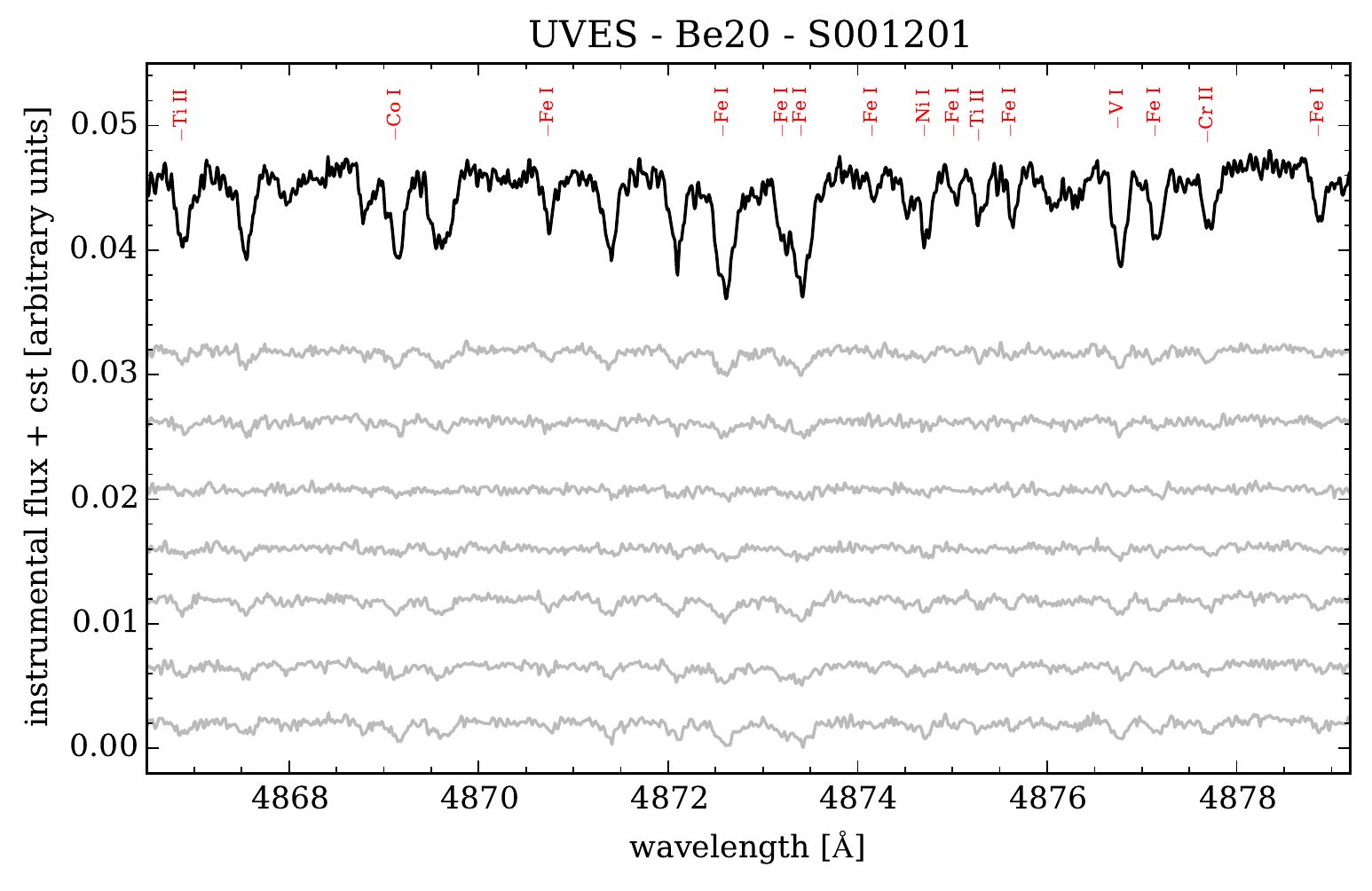}} \caption{\label{fig:spectrumBe20} Fragment of spectrum for a star of Be~20, highlighting some identified spectral lines. The grey spectra are the seven individual exposures. The black spectrum is the combined (co-added) exposure from which equivalent widths were measured.} \end{center}
\end{figure}

The data reduction was performed with the Reflex pipeline\footnote{\url{http://www.eso.org/sci/software/reflex/}}. The spectra were then corrected for heliocentric velocity, and multiple exposures were co-added to obtain a better signal-to-noise ratio (see Fig.~\ref{fig:spectrumBe20}), using the same procedure as adopted for the HIRES spectra.

\section{Spectroscopic analysis}
All spectra, whether they were acquired with HIRES or UVES, were analysed following the same procedure. 

We measured equivalent widths in the combined spectra with DOOp \citep{CantatGaudin14doop}, a pipeline based on DAOSPEC \citep{Stetson08} and originally designed to deal with the large number of spectra acquired by GES. The atomic data we used are those prepared for GES \citep{Heiter15} based on the VALD3 data \citep{Ryabchikova11}.

Stellar parameters and individual elemental abundances were obtained with FAMA \citep{Magrini13fama}, an automatic code based on Moog \citep{Sneden12}, making use of MARCS model atmospheres \citep[][]{Gustafsson08}. FAMA uses the classical equivalent widths method to determine the effective temperature by obtaining excitation equilibrium, surface gravity by finding the ionisation equilibrium, and microturbulence by eliminating abundance trends with equivalent width. We typically used 70 FeI lines and 8 FeII lines (see Table~\ref{tab:stellarparams}) for the determination of stellar parameters.

A certain number of spectra have a low signal-to-noise ratio. Given the large spectral range covered by the observation and the fact that the targets are cool, metal-rich stars, we observe a sufficient number of Fe lines to derive stellar parameters and [Fe/H] abundances from the equivalent widths method for all spectra. The abundances of elements such as Mg, for which fewer spectral lines are visible, are more affected by noise as their determination relies on poorer statistics.

The stellar parameters and individual elemental abundances are presented in Table~\ref{tab:stellarparams}. The abundances that we present are solar-scaled values \mbox{[X/H]=log($N_X$/$N_H$)$-$log($N_X$/$N_H$)$_{\odot}$}, where the reference solar values log($N_X$/$N_H$)$_{\odot}$ were taken from \citet{Grevesse07}\footnote{log($N_{Mg}$/$N_H$)$_{\odot}$=7.53; log($N_{Si}$/$N_H$)$_{\odot}$=7.51; log($N_{Ca}$/$N_H$)$_{\odot}$=6.31; log($N_{Ti}$/$N_H$)$_{\odot}$=4.90.}. The measured equivalent widths are available as an electronic table.

\begin{table*}
\begin{center}
	\caption{ \label{tab:stellarparams} Stellar parameters and elemental abundances for all stars in this study.}
	\small\addtolength{\tabcolsep}{-3pt}
	\begin{tabular}{c c c c c c | c c c | c c c | c c c | c c c | c c c}
\hline
\hline
oc & star & M & $\Teff$  & $\logg$ & v$_{mic}$ & \multicolumn{3}{c}{[Fe/H]} & \multicolumn{3}{c}{[Mg/H]} & \multicolumn{3}{c}{[Si/H]} & \multicolumn{3}{c}{[Ca/H]} & \multicolumn{3}{c}{[Ti/H]} \\
   &      &   & [K]      &         & [km\,s$^{-1}$] &   $\mu$ & $\sigma$ & n  & $\mu$ & $\sigma$ & n       & $\mu$ & $\sigma$ & n       & $\mu$ & $\sigma$ & n       & $\mu$ & $\sigma$ & n \\

\hline
Be~20 & S001201 & Y & 4733 $\pm$ 107 & 2.56 $\pm$ 0.21 & 1.29 $\pm$ 0.05       & -0.38 & 0.09 & 73   & 0.28 & 0.05 & 2   & 0.11 & 0.02 & 5   & 0.07 & 0.08 & 14   & 0.08 & 0.08 & 46   \\  
Be~20 & S001240 & Y & 4319 $\pm$ 66 & 1.55 $\pm$ 0.16 & 1.35 $\pm$ 0.05       & -0.50 & 0.07 & 56   & 0.28 & 0.01 & 2   & 0.06 & 0.07 & 5   & -0.09 & 0.04 & 4   & -0.10 & 0.05 & 39   \\  
Be~20 & S001403 & N & 4435 $\pm$ 121 & 4.95 $\pm$ 0.20 & 1.12 $\pm$ 0.05       & -0.02 & 0.09 & 57   & 0.44 & 0.09 & 2   & 0.15 & 0.03 & 5   & -0.32 & 0.11 & 4   & -0.11 & 0.10 & 46   \\  
Be~20 & S001519 & N & 4829 $\pm$ 118 & 3.90 $\pm$ 0.05 & 2.04 $\pm$ 0.09       & -0.44 & 0.13 & 54   & 0.70 & 0.00 & 1   & 0.18 & 0.11 & 5   & 0.19 & 0.07 & 2   & 0.30 & 0.16 & 39   \\  
Be~20 & S001666 & N & 4669 $\pm$ 105 & 4.73 $\pm$ 0.10 & 1.10 $\pm$ 0.05       & -0.45 & 0.09 & 77   & \multicolumn{2}{c}{...} & 0   & 0.12 & 0.06 & 4   & -0.05 & 0.06 & 7   & 0.14 & 0.10 & 52   \\  
Be~20 & S001716 & N & 4357 $\pm$ 74 & 1.67 $\pm$ 0.20 & 1.51 $\pm$ 0.05       & -0.68 & 0.07 & 62   & 0.41 & 0.01 & 2   & 0.14 & 0.04 & 6   & 0.09 & 0.06 & 7   & 0.01 & 0.05 & 39   \\  
\hline
Be~22 & K400 & Y & 4781 $\pm$ 91 & 2.81 $\pm$ 0.07 & 1.26 $\pm$ 0.05       & -0.30 & 0.13 & 95   & 0.11 & 0.02 & 2   & 0.10 & 0.06 & 10   & 0.00 & 0.04 & 10   & 0.08 & 0.06 & 34   \\  
Be~22 & K414 & Y & 4393 $\pm$ 97 & 2.09 $\pm$ 0.20 & 1.58 $\pm$ 0.05       & -0.43 & 0.11 & 71   & 0.41 & 0.00 & 2   & 0.15 & 0.01 & 5   & 0.05 & 0.05 & 5   & 0.09 & 0.07 & 44   \\  
Be~22 & K579 & Y & 4767 $\pm$ 96 & 2.51 $\pm$ 0.11 & 1.27 $\pm$ 0.05       & -0.34 & 0.13 & 93   & 0.13 & 0.02 & 2   & 0.10 & 0.09 & 10   & 0.08 & 0.03 & 8   & 0.07 & 0.07 & 30   \\  
\hline
Be~29 & K801 & Y & 4985 $\pm$ 91 & 2.55 $\pm$ 0.16 & 1.62 $\pm$ 0.03       & -0.51 & 0.09 & 62   & \multicolumn{2}{c}{...} & 0   & 0.15 & 0.06 & 3   & 0.09 & 0.06 & 8   & 0.09 & 0.04 & 10   \\  
Be~29 & K1032 & Y & 4991 $\pm$ 85 & 2.29 $\pm$ 0.17 & 1.58 $\pm$ 0.03       & -0.54 & 0.08 & 63   & \multicolumn{2}{c}{...} & 0   & 0.05 & 0.05 & 2   & 0.12 & 0.10 & 8   & 0.08 & 0.05 & 8   \\  

Be~29 & T159 & Y & 4995 $\pm$ 88 & 2.59 $\pm$ 0.17 & 1.60 $\pm$ 0.01       & -0.48 & 0.09 & 83   & 0.29 & 0.22 & 2   & 0.03 & 0.06 & 4   & 0.10 & 0.07 & 12   & 0.11 & 0.08 & 41   \\  
Be~29 & T257 & Y & 4871 $\pm$ 76 & 1.99 $\pm$ 0.04 & 1.52 $\pm$ 0.05       & -0.64 & 0.13 & 106   & 0.49 & 0.03 & 2   & 0.00 & 0.06 & 4   & 0.11 & 0.09 & 15   & 0.09 & 0.08 & 39   \\  
Be~29 & T398 & Y & 4965 $\pm$ 97 & 2.49 $\pm$ 0.20 & 1.56 $\pm$ 0.03       & -0.55 & 0.10 & 78   & 0.40 & 0.20 & 2   & 0.11 & 0.09 & 5   & 0.16 & 0.06 & 11   & 0.11 & 0.08 & 41   \\  
Be~29 & T602 & Y & 4949 $\pm$ 98 & 2.13 $\pm$ 0.14 & 1.65 $\pm$ 0.02       & -0.56 & 0.09 & 80   & 0.29 & 0.00 & 2   & 0.04 & 0.06 & 4   & 0.10 & 0.10 & 13   & 0.11 & 0.06 & 39   \\  
Be~29 & T933 & Y & 4981 $\pm$ 103 & 2.57 $\pm$ 0.19 & 1.61 $\pm$ 0.01       & -0.45 & 0.08 & 75   & 0.21 & 0.04 & 2   & -0.01 & 0.08 & 4   & 0.05 & 0.07 & 12   & 0.11 & 0.07 & 42   \\  
Be~29 & T1024 & Y & 3887 $\pm$ 37 & 1.15 $\pm$ 0.20 & 1.44 $\pm$ 0.01       & -0.46 & 0.11 & 49   & 0.25 & 0.03 & 2   & 0.12 & 0.12 & 5   & -0.09 & 0.02 & 2   & 0.08 & 0.12 & 27   \\

\hline
Be~66 & PJ785 & Y & 4769 $\pm$ 97 & 2.51 $\pm$ 0.20 & 1.15 $\pm$ 0.03       & -0.10 & 0.12 & 37   & -0.01 & 0.06 & 2   & 0.05 & 0.07 & 2   & 0.05 & 0.01 & 2   & -0.17 & 0.05 & 5   \\  
Be~66 & PJ934 & Y & 4937 $\pm$ 121 & 2.88 $\pm$ 0.18 & 1.97 $\pm$ 0.06       & -0.32 & 0.13 & 76   & 0.12 & 0.16 & 2   & 0.09 & 0.11 & 10   & 0.02 & 0.09 & 7   & 0.19 & 0.09 & 29   \\  

\hline
Be~73 & CC12 & N & 5001 $\pm$ 122 & 2.51 $\pm$ 0.17 & 1.52 $\pm$ 0.01       & -0.45 & 0.11 & 257   & 0.26 & 0.09 & 3   & 0.03 & 0.06 & 7   & 0.05 & 0.09 & 20   & 0.04 & 0.07 & 52   \\  
Be~73 & CC13 & N & 5679 $\pm$ 108 & 4.20 $\pm$ 0.04 & 1.04 $\pm$ 0.05       & 0.15 & 0.11 & 276   & 0.19 & 0.11 & 3   & -0.16 & 0.04 & 9   & -0.01 & 0.10 & 19   & -0.01 & 0.10 & 45   \\  
Be~73 & CC14 & Y & 4873 $\pm$ 127 & 2.89 $\pm$ 0.17 & 1.57 $\pm$ 0.05       & -0.23 & 0.11 & 149   & -0.14 & 0.00 & 1   & 0.02 & 0.03 & 3   & -0.19 & 0.03 & 3   & 0.10 & 0.09 & 35   \\  
Be~73 & CC15 & Y & 5093 $\pm$ 97 & 3.10 $\pm$ 0.11 & 1.15 $\pm$ 0.04       & -0.33 & 0.13 & 288   & 0.14 & 0.03 & 2   & -0.06 & 0.03 & 6   & 0.03 & 0.09 & 16   & 0.01 & 0.10 & 47   \\  
Be~73 & CC18 & N & 4961 $\pm$ 87 & 2.81 $\pm$ 0.10 & 1.41 $\pm$ 0.05       & -0.22 & 0.11 & 239   & 0.14 & 0.02 & 2   & -0.04 & 0.06 & 7   & 0.00 & 0.07 & 14   & 0.03 & 0.08 & 53   \\  
Be~73 & CC19 & N & 5965 $\pm$ 121 & 4.09 $\pm$ 0.04 & 1.22 $\pm$ 0.02       & 0.14 & 0.12 & 268   & 0.09 & 0.05 & 3   & -0.18 & 0.10 & 9   & -0.03 & 0.13 & 18   & -0.03 & 0.17 & 45   \\  
\hline
Be~75 & CC9 & Y & 4955 $\pm$ 116 & 2.51 $\pm$ 0.13 & 1.21 $\pm$ 0.03       & -0.38 & 0.14 & 89   & 0.52 & 0.06 & 2   & 0.09 & 0.03 & 5   & 0.18 & 0.10 & 13   & 0.17 & 0.11 & 48   \\  
Be~75 & CC22 & N & 5357 $\pm$ 60 & 3.43 $\pm$ 0.14 & 1.65 $\pm$ 0.07       & -0.26 & 0.10 & 43   & -0.01 & 0.00 & 1   & -0.03 & 0.03 & 4   & -0.02 & 0.06 & 5   & 0.24 & 0.09 & 27   \\  
\hline
Rup~4 & CC18 & N & 5001 $\pm$ 77 & 3.05 $\pm$ 0.09 & 1.19 $\pm$ 0.04       & -0.33 & 0.12 & 272   & 0.24 & 0.09 & 2   & -0.02 & 0.05 & 7   & 0.05 & 0.10 & 21   & 0.06 & 0.08 & 49   \\  
Rup~4 & CC29 & N & 4975 $\pm$ 121 & 2.94 $\pm$ 0.13 & 1.44 $\pm$ 0.02       & -0.28 & 0.12 & 244   & 0.25 & 0.07 & 2   & -0.07 & 0.06 & 8   & 0.00 & 0.10 & 14   & 0.13 & 0.09 & 54   \\  
Rup~4 & CC3 & Y & 5153 $\pm$ 104 & 2.70 $\pm$ 0.15 & 1.71 $\pm$ 0.01       & -0.15 & 0.09 & 207   & 0.26 & 0.01 & 2   & -0.02 & 0.08 & 8   & 0.00 & 0.08 & 12   & -0.02 & 0.09 & 51   \\  
Rup~4 & CC4 & Y & 5079 $\pm$ 99 & 2.48 $\pm$ 0.11 & 1.57 $\pm$ 0.02       & -0.09 & 0.09 & 203   & 0.21 & 0.04 & 3   & -0.07 & 0.07 & 9   & -0.01 & 0.07 & 14   & -0.07 & 0.10 & 47   \\  
Rup~4 & CC8 & Y & 5155 $\pm$ 105 & 2.66 $\pm$ 0.10 & 1.38 $\pm$ 0.02       & -0.19 & 0.11 & 264   & 0.22 & 0.03 & 3   & -0.10 & 0.06 & 8   & -0.01 & 0.06 & 15   & -0.04 & 0.09 & 47   \\  
\hline
Rup~7 & CC2 & Y & 5117 $\pm$ 79 & 2.01 $\pm$ 0.07 & 1.39 $\pm$ 0.02       & -0.39 & 0.11 & 83   & -0.16 & 0.00 & 1   & -0.03 & 0.10 & 6   & 0.09 & 0.13 & 14   & -0.02 & 0.09 & 28   \\  
Rup~7 & CC4 & Y & 5027 $\pm$ 70 & 2.07 $\pm$ 0.10 & 1.91 $\pm$ 0.04       & -0.33 & 0.07 & 56   & 0.09 & 0.19 & 2   & 0.04 & 0.08 & 5   & 0.04 & 0.07 & 9   & -0.02 & 0.08 & 34   \\  
Rup~7 & CC5 & Y & 5151 $\pm$ 79 & 2.23 $\pm$ 0.05 & 2.39 $\pm$ 0.05       & -0.43 & 0.08 & 66   & 0.10 & 0.32 & 2   & 0.04 & 0.07 & 5   & 0.02 & 0.05 & 8   & 0.05 & 0.10 & 28   \\  
Rup~7 & CC6 & Y & 5037 $\pm$ 68 & 1.83 $\pm$ 0.06 & 1.81 $\pm$ 0.02       & -0.34 & 0.08 & 59   & 0.08 & 0.25 & 2   & -0.04 & 0.05 & 5   & 0.09 & 0.05 & 8   & -0.04 & 0.09 & 28   \\  
Rup~7 & CC7 & Y & 5015 $\pm$ 72 & 2.22 $\pm$ 0.06 & 1.83 $\pm$ 0.04       & -0.39 & 0.11 & 65   & -0.10 & 0.00 & 1   & 0.13 & 0.06 & 3   & 0.05 & 0.07 & 9   & 0.03 & 0.11 & 30   \\  

\hline
Sau~1 & C91 & Y & 5138 $\pm$ 106 & 2.57 $\pm$ 0.16 & 1.68 $\pm$ 0.02       & -0.37 & 0.08 & 96   & \multicolumn{2}{c}{...} & 0  & 0.11 & 0.04 & 5   & 0.02 & 0.11 & 12   & 0.09 & 0.03 & 9   \\  
Sau~1 & C122 & Y & 4900 $\pm$ 65 & 2.58 $\pm$ 0.07 & 1.43 $\pm$ 0.01       & -0.34 & 0.09 & 95   & \multicolumn{2}{c}{...} & 0   & 0.14 & 0.06 & 2   & 0.02 & 0.07 & 7   & 0.08 & 0.04 & 13   \\

\hline
Tom~2 & P146 & Y & 5271 $\pm$ 129 & 3.37 $\pm$ 0.10 & 2.24 $\pm$ 0.04       & -0.21 & 0.17 & 78   & 0.75 & 0.24 & 2   & 0.08 & 0.10 & 6   & -0.07 & 0.13 & 10   & 0.30 & 0.14 & 49   \\  
Tom~2 & P158 & Y & 5251 $\pm$ 94 & 3.26 $\pm$ 0.20 & 2.46 $\pm$ 0.01       & -0.34 & 0.20 & 72   & 0.28 & 0.39 & 2   & 0.16 & 0.15 & 6   & 0.03 & 0.11 & 11   & 0.35 & 0.16 & 44   \\  
Tom~2 & P162 & N & 5521 $\pm$ 142 & 3.70 $\pm$ 0.06 & 1.33 $\pm$ 0.03       & 0.11 & 0.17 & 82   & -0.14 & 0.25 & 2   & -0.03 & 0.07 & 5   & -0.29 & 0.16 & 9   & 0.35 & 0.21 & 49   \\  
Tom~2 & P164 & N & 4973 $\pm$ 100 & 3.04 $\pm$ 0.15 & 1.64 $\pm$ 0.01       & -0.48 & 0.13 & 81   & 0.33 & 0.26 & 2   & 0.23 & 0.09 & 6   & 0.12 & 0.06 & 10   & 0.25 & 0.10 & 47   \\  
Tom~2 & P165 & N & 5195 $\pm$ 122 & 2.96 $\pm$ 0.12 & 1.29 $\pm$ 0.01       & 0.06 & 0.14 & 65   & -0.19 & 0.14 & 2   & 0.01 & 0.10 & 4   & 0.02 & 0.12 & 9   & 0.16 & 0.17 & 42   \\  

\hline

\end{tabular}
\tablefoot{M indicates whether a star is considered a cluster member (Y) or not (N).
\newline For each element we list the mean abundance ($\mu$), standard deviation ($\sigma$), and number of lines used (n).}
\end{center}
\end{table*}

\subsection{Results for individual clusters} \label{sec:results}
The average [Fe/H] was computed as a weighted mean considering the uncertainty on the individual iron abundance of each member star. 
The uncertainty on the mean [Fe/H] was computed as the statistical error of the average (mean error divided by the square root of the number of member stars).

We computed $\alpha$-abundances as the average of Mg, Si, Ca, and Ti abundances.

\subsubsection{Berkeley 20}
Following \citet{Sestito08}, we consider the stars S001201 and S001240 (with radial velocities of 79\,km\,s$^{-1}$) to be likely cluster members. Three stars exhibit dwarf-like surface gravities and must therefore be foreground field stars. A fourth star (SS001716) appears to be a cool giant, but its radial velocity of 38\,km\,s$^{-1}$ is incompatible with the cluster (see Fig.\ref{fig:summary_part1}).
We obtain a mean [Fe/H] of -0.42$\pm$0.05\,dex. This value sits between the results of S08 (-0.3$\pm$0.02) and those of Y05 (-0.49$\pm$0.06), although it is closer to the result of Y05, even though the data analysed here are the spectra acquired by S08.

For both member stars (S001201 and S001240) we obtain very similar parameters to those obtained by S08. The lower Fe abundance we obtain can be due to a different approach to normalising the spectrum and measuring equivalent widths \citep[S08 used IRAF and SPECTRE][]{Fitzpatrick87}, as well as a different choice of model atmospheres \citep[S08 used the models of][]{Kurucz93}.

\subsubsection{Berkeley 22}
All three targets of Be~22 show consistent radial velocities (RVs) and metallicities (Fig.\ref{fig:summary_part1}). We therefore consider them cluster members. We obtain a [Fe/H] value of -0.36$\pm$0.07, which is in good agreement with V05.

\subsubsection{Berkeley 29}
All seven targets in Be~29 show consistent RVs and metallicities (see Fig.\ref{fig:summary_part1}). We find a mean [Fe/H] of -0.52$\pm$0.03, in good agreement with C04, but slightly more metal poor than the result of S08. We note that the mean [Fe/H] derived from the two HIRES spectra is -0.53, which is almost exactly the average of the whole sample. 

One star was observed both with HIRES (under the name K801) and UVES (T398). We conducted the analysis of those spectra independently. The stellar parameters ($\Teff$, $\logg$) and iron abundance found in both cases are in remarkable agreement, differing by only 20\,K, 0.06, and 0.03\,dex (respectively). The abundances of $\alpha$-elements differ on average by 0.04\,dex.

\subsubsection{Berkeley 66}
Both spectroscopic targets of Be~66 show consistent RVs and metallicities (see Fig.\ref{fig:summary_part1}). We find a metallicity of -0.21$\pm$0.09, which is significantly higher than the value of -0.48$\pm$0.24 found by V05. The stellar parameters adopted by V05 for the star 785, which is the only star for which they derived abundances, are very similar to those we derived here. They provide equivalent widths (measured with IRAF's \textit{splot} task) for 32 lines, out of which only 8 are in common with our line selection, and for which the equivalent widths measured by us are on average about 20\,m$\AA$ larger. We therefore attribute the difference in iron abundance to the equivalent width measurement procedure.

\subsubsection{Berkeley 73}
Three targets in Be~73 exhibit similar RVs around 74\,km\,s$^{-1}$. One of these targets (CC13) happens to be a dwarf and shows a discrepant metallicity (see Fig.\ref{fig:summary_part1}). We therefore only consider CC14 and CC15 as likely cluster members. The mean metallicity of those two stars is -0.28$\pm$0.08, which is in agreement with the value -0.22$\pm$0.10 found by C07. 

\subsubsection{Berkeley 75}
For both targets, we obtain very similar parameters to those found by C07. These stars exhibit discrepant metallicities (-0.38$\pm$0.14 and -0.27$\pm$0.10 for CC9 and CC22, respectively) and RVs (75 and 95\,km\,s$^{-1}$, respectively). A PARSEC isochrone \citep{Bressan12} for the age of Be~75 (4\,Gyr) shows the expected surface gravity is $\logg=2.5$ and the effective temperature $\Teff=4800$\,K. The surface gravity $\logg=3.4$ we find for CC22 is not compatible with it being a red clump star of Rup~4. We therefore consider that the star CC9 is a probable cluster member. This choice is the opposite of what was done by C07, who found a metallicity of -0.22$\pm$0.20 for C22 and consider it was the most likely cluster member.

\subsubsection{Ruprecht 4}
The targets for this cluster separate into two groups: one with mean metallicity $\sim$-0.15 around RV=45\,km\,s$^{-1}$, and one with mean metallicity $\sim$-0.3 around RV=62\,km\,s$^{-1}$ (see Fig.\ref{fig:summary_part2}).  Those two groups are also seen to separate in the $\Teff$ -- $\logg$ plane, with CC3, CC4, and CC8 showing higher temperatures and lower surface gravities than CC18 and CC29 (bottom left panel of Fig.\ref{fig:summary_part2}). A PARSEC isochrone for the age of Rup~4 (0.8\,Gyr) shows that red clump stars are not expected to exhibit surface gravities higher than $\logg=2.8$. Their effective temperature is expected to be 5100\,K for slightly subsolar metallicities, or above 5100\,K if more metal poor, which rules out CC18 and CC29 as possible cluster members.

We therefore adopt the same choice as C07, and consider CC3, CC4, and CC8 as likely cluster members. Their average metallicity is -0.14$\pm$0.06, which is in agreement with the result of -0.09$\pm$0.05 obtained by C07.

\subsubsection{Ruprecht 7}
All seven targets of Rup~7 show consistent RVs and metallicities (see Fig.\ref{fig:summary_part2}), and can be considered likely cluster members. Their mean metallicity is -0.37$\pm$0.04, which is slightly more metal poor than the value of -0.26$\pm$0.05 of C07.

\subsubsection{Saurer 1}
Both spectroscopic targets of Sau~1 show consistent RVs and metallicities (see Fig.\ref{fig:summary_part2}). We obtain a mean metallicity of -0.36$\pm$0.06 for those two stars, which is in very good agreement with the value of -0.38$\pm$0.14 derived by C04.

\subsubsection{Tombaugh 2}
Three of the five targets in Tom~2 exhibit similar RVs around 122\,km\,s$^{-1}$ (stars P146, P158, and P162). One of those three targets shows a significantly higher iron abundance (see Fig.\ref{fig:summary_part2}). Under the assumption that the cluster must be chemically homogeneous, we only consider the stars P146 and P158 as cluster members. We compute a mean metallicity of -0.27$\pm$0.13, in agreement with previous results (see references for Tom~2 in Sect.~\ref{sec:introclusters}). Tom~2 is the cluster for which we find the highest $\alpha$-abundance in the whole sample, but it also exhibits the largest uncertainty on this value. The observations for this cluster happen to have the lowest S/N ratio in the sample (see Table~\ref{tab:targets}).

\section{Kinematics and orbit integration} \label{sec:orbits}
When tracing the Galactic metallicity gradient, studies typically make use of the present-day Galactocentric radius of the tracers (here, open clusters), however, a correct interpretation of the metallicity gradient should not overlook the dynamics of the tracers involved. The present-day Galactocentric radius of a cluster may be different from its birth radius, and the tracer may represent a population originating from a different location in the disk of the Milky Way. A cluster can travel inwards or outwards from its birthplace because of radial migration, which is a change in the guiding radius of the orbit caused by resonant interactions with non-axisymmetrical features such as a bar or spiral arms \citep[][]{Sellwood02,Roskar08,Minchev14} or because it follows an eccentric orbit. Systematic studies of the orbits of the Galactic cluster population include \citet{Wu09}, \citet{Magrini10}, and \citet[][hereafter VdP10]{VandePutte10}. We only found orbit reconstructions in the literature for Be~29 (VdP10) and Be~20 \citep[VdP10 and][]{Wu09}.

Reconstructing the orbit of an object requires knowing six parameters: its 3D position (e.g. sky coordinates ($\alpha$,$\delta$) and heliocentric distance d$_{\odot}$) and its 3D velocity (e.g. proper motions ($\mu_{\alpha}$,$\mu_{\delta}$ and radial velocity RV). Assuming a certain gravitational potential for the Milky Way, the orbit can be integrated forwards, predicting the position of the object at a future time, or backwards, reconstructing the position at a past epoch. We have performed orbit integrations with the publicly available package {\tt galpy} \citep{Bovy15}, making use of the static, axisymmetric (does not feature the Galactic bar) potential {\tt MWPotential2014} shipped with the code. We assumed R$_{\mathrm{GC,\odot}}$=8.3\,kpc \citep{Eisenhauer05} and $v_{\mathrm{LSR}}$=240\,km\,s$^{-1}$.

The proper motions we used for each cluster are those listed in the MWSC catalogue \citep{Kharchenko13}, which are based on the PPMXL catalogue \citep{Roser10}. To assess the impact of the uncertainty on the proper motion, 2000 orbits were simulated in each case, adding Gaussian random errors corresponding to the proper motions uncertainty listed in MWSC. We also added Gaussian errors corresponding to a 5\% uncertainty on d$_{\odot}$ and 10\% uncertainty on the cluster age. The other parameters ($\alpha$,$\delta$, and RV) were kept fixed to the values listed in Table~\ref{tab:outerclusters}. We integrated back in time, for a total duration corresponding to the age of the cluster.  We found no proper motion values in literature for the cluster Rup~7 and could not perform orbit reconstruction for this object.

\subsection{Reconstructed orbital parameters}
\begin{figure}[ht]
\begin{center} \resizebox{\hsize}{!}{\includegraphics[scale=0.8]{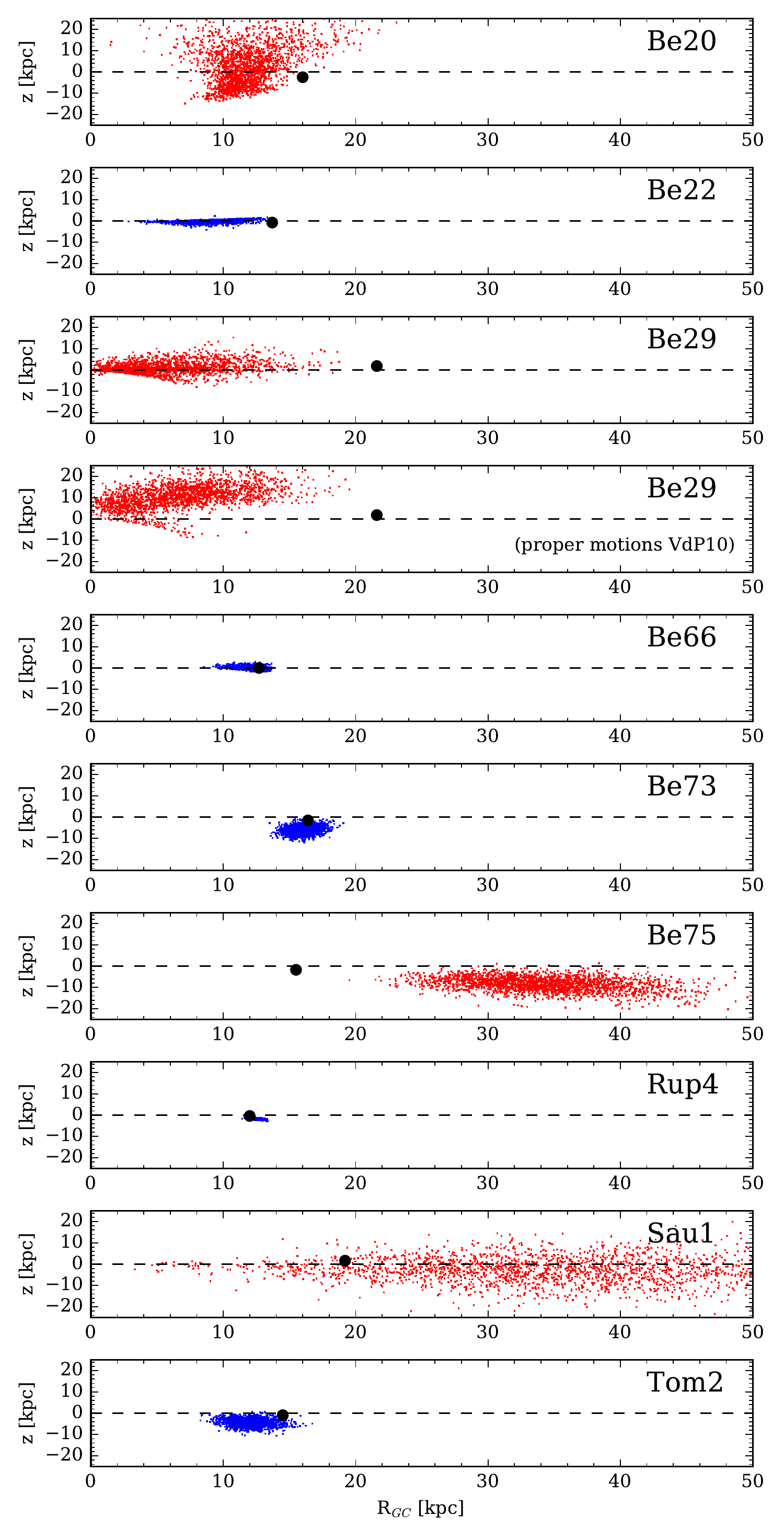}} \caption{\label{fig:Rbirth_zbirth} Reconstructed positions at time of birth, obtained integrating the current orbit back in time over a period corresponding to their age, for each cluster (coloured points). The black dot indicates the present position. The proper motions used are those of MWSC unless specified. The cases for which we find unexpectedly large dispersions are shown in red.} \end{center}
\end{figure}

The main parameters of the reconstructed orbits are listed in Table~\ref{tab:orbitparams}. By integrating the orbit back to a time corresponding to the age of the clusters, we evaluate their location at the moment of their birth. The positions recovered in the 2000 runs performed for each cluster are shown in Fig.~\ref{fig:Rbirth_zbirth}.

\begin{table*}
\begin{center}
	\caption{ \label{tab:orbitparams} Parameters for the reconstructed orbits.}
	\small\addtolength{\tabcolsep}{-1pt}
	\begin{tabular}{c c c c c c c c c c c c}
	\hline
	\hline
	OC & $\alpha$ & $\delta$ & $\mu_{\alpha}$ cos$\delta$ & $\mu_{\delta}$ & PM error & R.V. & R$_{\mathrm{GC}}$ & z & R$_{\mathrm{birth}}$ & z$_{\mathrm{birth}}$ & E\\
	   & (J2000) & (J2000) & [mas/yr] & [mas/yr] & [mas/yr] & [km\,s$^{-1}$] & [kpc] & [kpc] & [kpc] & [kpc] & [$\times10^4$ kg$^2$s$^{-2}$] \\
	\hline
Be~20  & 83.250 & 0.217 & -1.6 & -4.3 & 0.9 & 	79.5 & 16.0 & -2.5 & 11.8 $\pm$ 2.3 		& 2.0 $\pm$ 8.9& -2.29 $\pm$ 0.7 \\ 
Be~20* & 83.250 & 0.217 & 1.5 & -4.1  & 0.9 & 	79.5 & 16.0 & -2.5 & 13.9 $\pm$ 2.1 		& -5.0 $\pm$ 4.3& -3.9 $\pm$ 0.3 \\ 
Be~22  & 89.600 & 7.833 & 1.8 & -3.9 & 0.7 & 	95.3 & 13.7 & -0.8 & 9.3 $\pm$ 1.8 		& 0.1 $\pm$ 0.7 & -4.52 $\pm$ 0.22 \\
Be~29  & 103.325 & 16.917 & 1.1 & -3.9 & 0.5 & 	24.8 & 21.6 & 1.9 & 5.9 $\pm$ 3.5 		& 0.5 $\pm$ 2.7 & -2.73 $\pm$ 0.24 \\
Be~29* & 103.325 & 16.917 & -0.1 & -4.8 & 0.5 & 	24.8 & 21.6 & 1.9 & 6.8 $\pm$ 3.7 		& 10.5 $\pm$ 5.3 & -2.04 $\pm$ 0.49 \\
Be~66 & 46.075  & 58.767 & 2.8 & 0.0 & 0.3 & 	-50.6 & 12.7 & 0.1 & 12.4 $\pm$ 0.7 		& -0.7 $\pm$ 0.8 & -4.11 $\pm$ 0.13 \\
Be~73 & 95.525  &  -6.317 & 3.5 & -1.7 & 0.7 &   95.7 & 16.4 & -1.6 & 16.1 $\pm$ 0.8 		& -5.9 $\pm$ 1.7 & -3.12 $\pm$ 0.4 \\
Be~75 & 102.246 & -23.992 & 0.5 & 5.3 & 0.7 & 	94.6 & 15.5 & -1.7 & 34.2 $\pm$ 4.9 		& -8.3 $\pm$ 3.2 & 2.07 $\pm$ 1.13 \\
Rup~4 & 102.225 & -10.533 & 1.8 & 2.2 & 0.4 & 	47.5 & 12.0 & -0.4 & 12.4 $\pm$ 0.3 		& -1.6 $\pm$ 0.2 & -3.01 $\pm$ 0.2 \\
Sau~1 & 109.575 & 1.887 & 3.5 & -6.7 & 0.8 & 	98.0 & 19.2 & 1.6 & 33.9 $\pm$ 10.9 		& -2.5 $\pm$ 6.1 & 0.27 $\pm$ 1.48 \\
Tom~2 & 105.771 & -20.817 & 1.7 & 2.0 & 0.7 & 	121.8 & 14.2 & -0.9 & 12.0 $\pm$ 1.1 		& -4.2 $\pm$ 1.6 & -2.61 $\pm$ 0.56 \\
	\hline
	\end{tabular}
\tablefoot{Proper motions from the MWSC catalogue \citep{Kharchenko13} unless specified.

* the second computation for Be~20 uses the proper motions of DAML02. The second computation for Be~29 was performed using the proper motions used by VdP10. }
\end{center}
\end{table*}

\begin{figure}[ht]
\begin{center} \resizebox{\hsize}{!}{\includegraphics[scale=0.8]{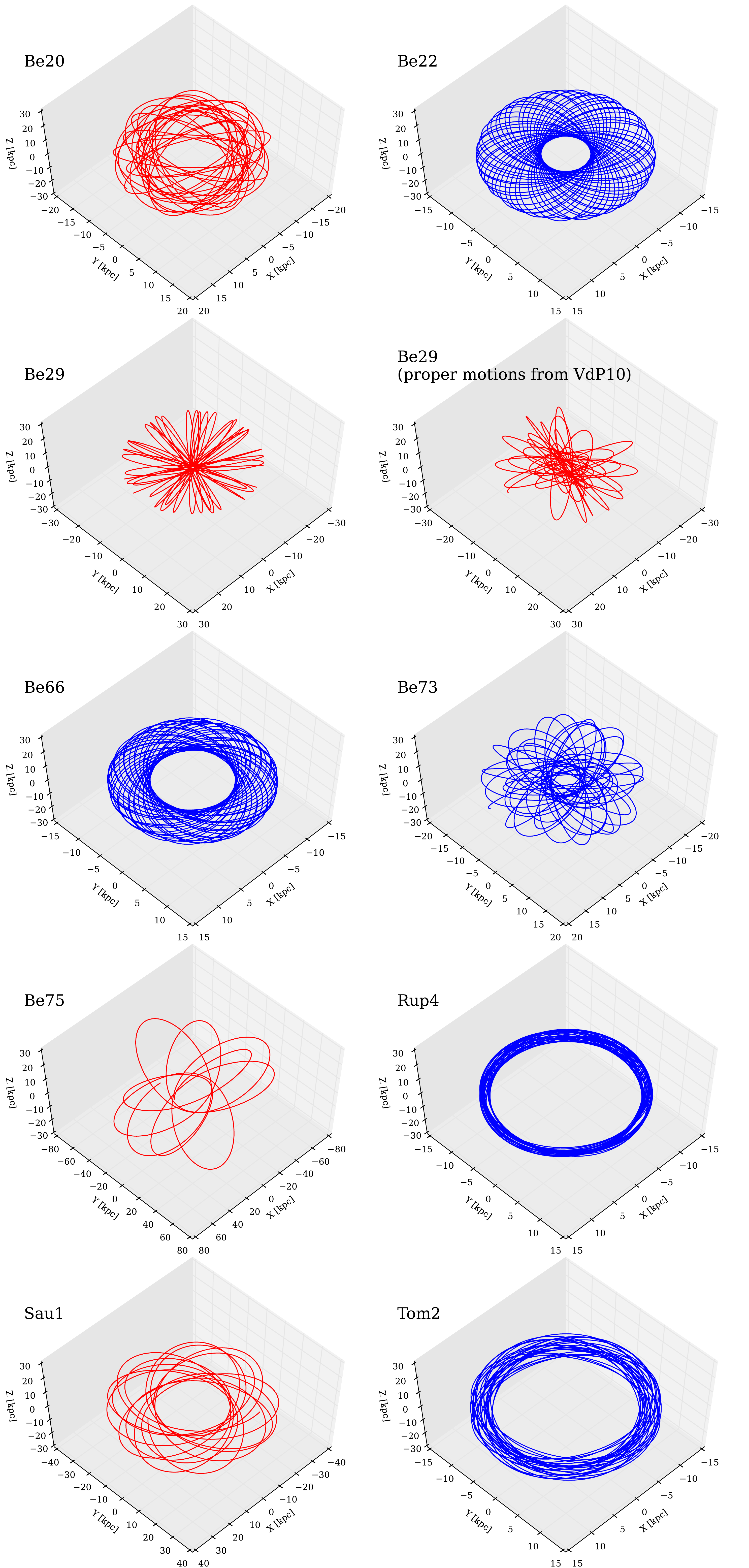}} \caption{\label{fig:3d_orbits} Integrated orbits for the clusters studied in this sample. Those four clusters for which we find irregular orbits are traced in red. The proper motions used are those of MWSC unless specified. } \end{center}
\end{figure}

The resulting orbits are shown in Fig.~\ref{fig:3d_orbits}. We notice that five clusters (namely Be~22, Be~66, Be~73, Rup~4 and Tom~2) follow regular crown orbits, while four (Be~20, Be~29, Be~75, and Sau~1) appear to follow more chaotic and unexpected trajectories, which are also characterised by a large dispersion in the reconstructed birthplaces (shown in Fig.~\ref{fig:Rbirth_zbirth}). 

The irregular orbit of Be~20 was previously reported by VdP10, who used similar proper motions ($\mu_{\alpha}$cos$\delta$=1.51\,mas.yr$^{-1}$, $\mu_{\delta}$=-4.11\,mas.yr$^{-1}$). They also report on the unexpected orbit of Be~29, in particular the fact that it appears to follow a retrograde (counter-rotating) orbit in most of their simulated runs. We reach a similar conclusion using the proper motions listed in MWSC. We also performed integrations using the proper motion values of VdP10 for Be~29, and obtained similar results. 

The distant OC Sau~1 appears to be on a retrograde orbit as well, and in fact does not appear to be gravitationally bound to the Milky Way (its orbit carries a total energy $E>0$). 

Finally, Be~75 follows a complex, but prograde, orbit, and appears unbound too. Since our radial velocity value for Be~75 comes from one single star with uncertain membership (see Sect.~\ref{sec:results}), we computed orbits varying the RV from -200 to +200\,km\,s$^{-1}$. These orbits correspond to an unbound object in all cases.

We do not suggest that Be~20, Be~29, Be~75, and Sau~1 actually did form at very large Galactocentric distances and/or far from the Galactic plane. The first possible explanation for those peculiar orbits may simply be the quality of the proper motion determination, considering the large distance to those objects and the fact that their brightest stars barely reach magnitude $V$=15. If we trust the proper motion values, we interpret those results as a hint that the orbits of these objects have been perturbed during their life, and we cannot reconstruct their history by simply integrating their current orbits back in time. In particular, Be~20 appears more metal poor than other clusters at the same Galactocentric radius (see Sect.~\ref{sec:gradient} and Fig.~\ref{fig:gradient_Both}), and may have formed in the outer parts of the disk. Similarly, the extremely large Galactocentric radius of Be~29 and Sau~1 could be due to perturbed orbits sending them outwards from their birth radius. 

A further argument in favour of these four clusters genuinely having peculiar orbits is their large distance from the Galactic plane, as they are in fact the four objects with the largest z in our sample. Since Galactic clusters are expected to be found close to the plane of the Milky Way, their location far from the plane indicates that they might have a peculiar orbital history, which would have caused the observed kinematics and large vertical excursion of their current orbit.

\begin{figure}[ht]
\begin{center} \resizebox{\hsize}{!}{\includegraphics[scale=0.8]{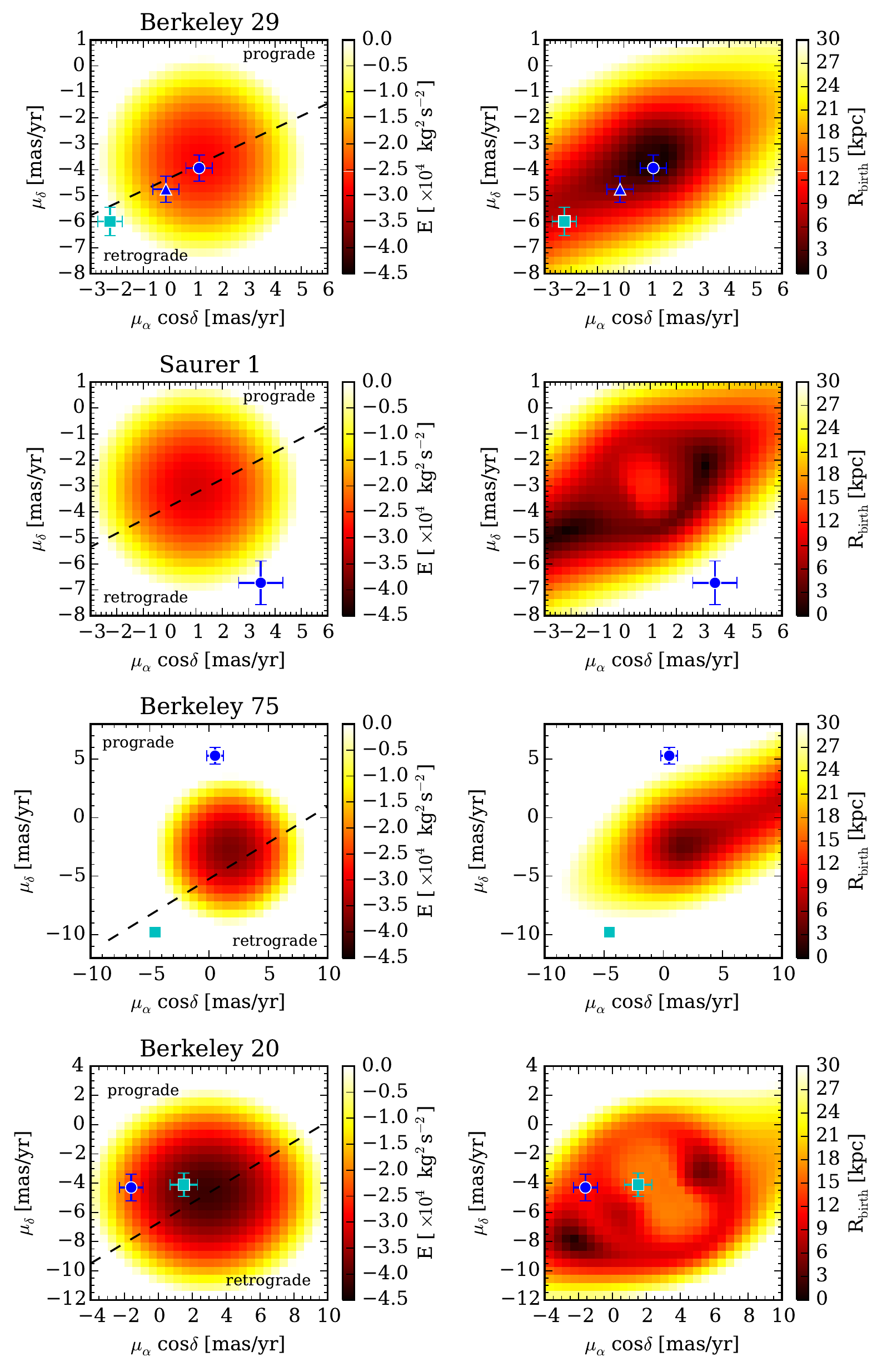}} \caption{\label{fig:E_rbirth_Be29_Sau1_Be75_Be20} \textit{Top left:} total energy for an object at the location and radial velocity of Be~29, as a function of its proper motions. The white area indicates combinations of proper motions for which $E>0$ (i.e. for which the object is not gravitationally bound to the Milky Way). The dashed line separates the regions for which the orbits are prograde and retrograde. \textit{Top right:} expected birth radius as a function of proper motions. The blue dot indicates the proper motions found in MWSC. The cyan squares indicate the proper motions of DAML02. The triangle indicates the proper motions of VdP10. \textit{Other rows:} same as previous, for Sau~1, Be~75, and Be~20.} \end{center}
\end{figure}

The most critical parameters in the reconstruction of the orbits of those distant objects are proper motions. We computed the parameters of a set of orbits varying the values of $\mu_{\alpha}$ and $\mu_{\delta}$ and keeping the other four parameters ($\alpha$, $\delta$, d$_{\odot}$, and RV) fixed to understand what the proper motions of those problematic objects should be for them to follow more regular orbits. The total energy (E) and Galactocentric radius at birth (R$_{\mathrm{GC}}$) obtained with various combinations of proper motions are shown in Fig.~\ref{fig:E_rbirth_Be29_Sau1_Be75_Be20}. For reference, the blue dots indicate the proper motions used in this study, while cyan squares show the values listed in the catalogue of \citet[][hereafter DAML02]{Dias02}.

For Be~29 we can see that the values quoted by MWSC, DAML02 and VdP10 are significantly different, but all of these values correspond to an elongated retrograde orbit. The dashed line in Fig.~\ref{fig:E_rbirth_Be29_Sau1_Be75_Be20} separates the prograde from retrograde orbits, and shows that within the formal uncertainty on its proper motions, it is also possible that Be~29 follows an elongated prograde orbit. This result is similar to the conclusion of \citet{VandePutte10}, who find retrograde orbits for Be~29 in 84\% of their simulations, and prograde orbits in the remaining cases.
All three values of proper motions also correspond to a birthplace in the inner disk. Given its low metallicity, it is unlikely that Be~29 was truly formed in the inner disk. A more likely scenario would be that this object formed in the outer disk and that its orbit was perturbed at some point during the 4.5\,Gyr of its life.

Applying the same diagnostics to Sau~1, we determine that its proper motions are only compatible with that of a retrograde orbit, and the total energy of its orbit is that of an unbound object. The location of Sau~1 at the time of birth (5\,Gyr ago) appears to be far out of the Galactic disk (see Fig.~\ref{fig:Rbirth_zbirth}). This would mean that either Sau~1 is of extragalactic origin, or that is orbit has been perturbed since its formation and it is now being ejected from our Galaxy. The catalogue of DAML02 gives \mbox{$\mu_{\alpha}$cos$\delta$=-6.62\,mas.yr$^{-1}$, $\mu_{\delta}$=8.13\,mas.yr$^{-1}$}, which is suspiciously high for such a distant object and would correspond to a certainly unbound trajectory.

Finally, Be~20 appears to be on a prograde bound trajectory whether we consider the proper motions of MWSC or DAML02. In both cases, its location 6\,Gyr ago (at the time of its birth) appears to be closer to the Galactic centre than its current location. This is unexpected, as Be~20 is more metal poor than other clusters at this Galactocentric radius (see Fig.~\ref{fig:gradient_Both}). It is therefore difficult to explain the present-day location of Be~20 with a simple elliptical orbit.

\subsection{Sources of error}
The determination of the mean proper motion of a cluster relies on the individual measurements of the proper motions for each star; for instance, errors in the PPMXL and UCAC4 catalogues are typically above 3\,mas.yr$^{-1}$ for stars fainter than $V$=15. However, this determination also depends on the methodology and the choice of cluster stars. Cluster membership is difficult to establish when only the brightest stars of a cluster have proper motion or radial velocity measurements. \citet{Dias14} compared the proper motions of clusters listed in various catalogues, and show that the values typically differ by more than the quoted formal errors. The scatter between the values obtained by different studies is expected to be especially large for distant clusters, as identifying cluster stars and matching them with proper motion catalogues may become a difficult task for faint objects. It is obvious from Fig.~\ref{fig:E_rbirth_Be29_Sau1_Be75_Be20} that different catalogues sometimes quote widely discrepant proper motions for a given object, and since both values listed for Be~75 correspond to chaotic orbits, we must consider the possibility that they are both incorrect. 

The upcoming catalogue of the astrometric Gaia mission \citet{Perryman01} will provide all-sky photometry and proper motions for stars down to $V$=20. The post-launch science performance assessments suggest the proper motion error at the faint end should be of the order of 70$\mu$mas for M stars. In the $V=15$--16 range, the Radial Velocity Spectrometer of Gaia will also provide radial velocities with a precision of 15\,km\,s$^{-1}$ for red stars. This precision is not sufficient for accurate kinematic studies and reliable membership determination, but makes it easier to select probable members and priority targets for high-resolution spectroscopic observations. Gaia is therefore expected to make a significant contribution to studies of the Galactic disk, and such data will allow us to confirm or disprove the peculiar orbits of Be~20, Be~29, Be~75, and Sau~1. 

The Galactic potential used in reconstructing particle orbits are traditionally static and axisymmetric, for instance \citet{Fellhauer06} \citep[used by][]{Law05,VandePutte10} or \citet{Allen91} \citep[used by][]{Wu09}. This study, making use of the model of \citet{Bovy15}, is no exception. Such potentials are unable to account for the mechanism of radial migration, that is, particles traveling inwards or outwards owing to a change in their guiding radius while staying on circular orbits, which is due to transient, non-axisymmetric patterns such as bars and spiral arms. This approach is therefore able to reveal whether the orbit of a tracer spans a wide range in Galactocentric radius because of its eccentricity, but is unable to reconstruct the original birthplace of a tracer if it underwent radial migration.

\section{Abundance gradients} \label{sec:gradient}
We investigated individual abundances of Fe, and of Mg, Si, Ca, and Ti (the so-called $\alpha$-elements). This is, to our knowledge, the first homogeneous analysis of a significant number of clusters in the Galactocentric radius range 12--20\,kpc.

\begin{figure*}[ht]
\begin{center} \resizebox{\hsize}{!}{\includegraphics[scale=0.8]{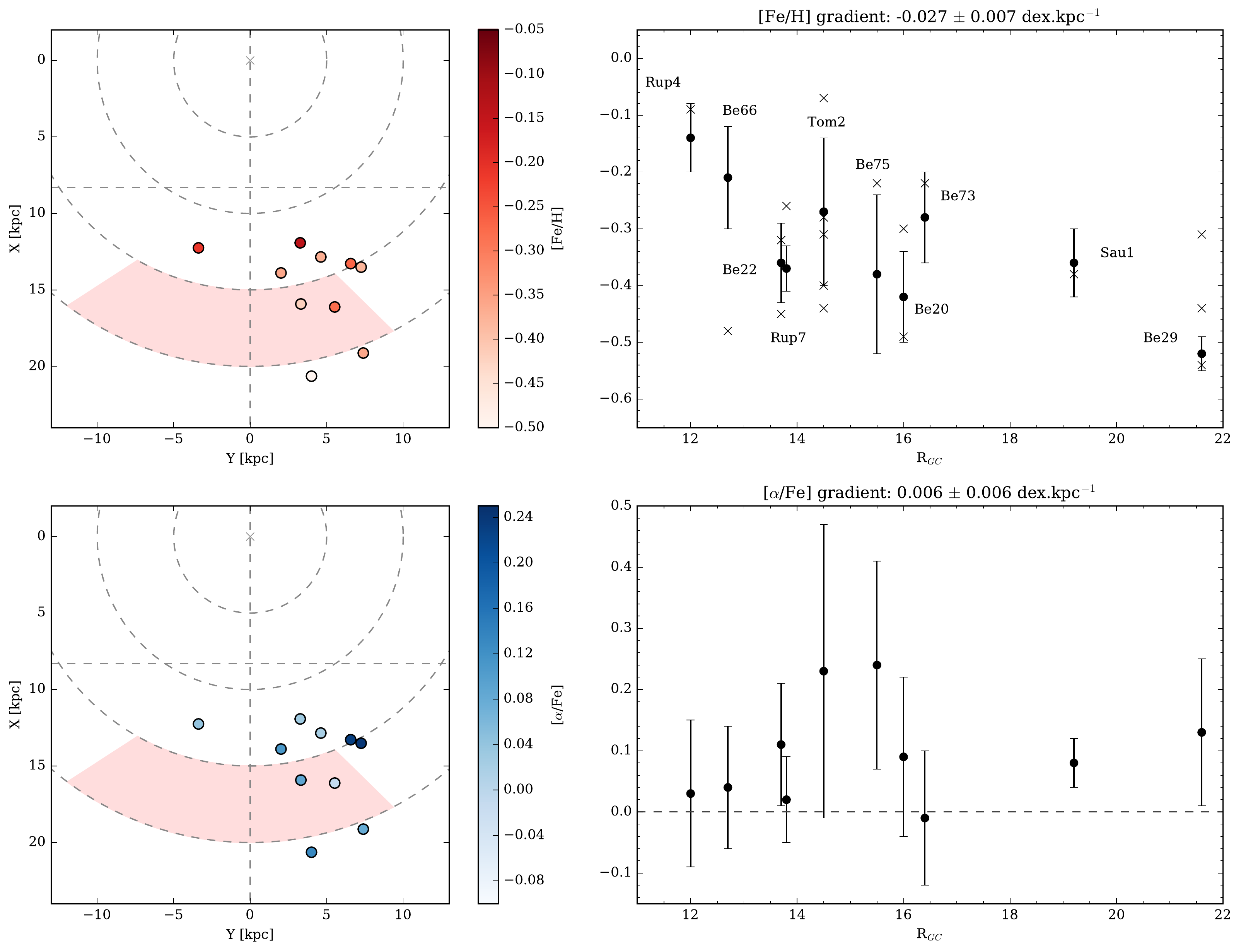}} \caption{\label{fig:gradient_Both} \textit{Top-left:} positions of the ten OCs under study, projected on the Galactic plane. The Sun is located at (X,Y)=(0,8.3). The shaded area indicates the Monoceros structure. \textit{Top-right:} [Fe/H] against Galactocentric radius for all ten clusters. The crosses indicate previous metallicity determinations from high-resolution spectroscopy (see references in Sect.~\ref{sec:introclusters}). \textit{Bottom row:} same as top row, for [$\alpha$/Fe] abundances (taken as the mean of Mg, Si, Ca, and Ti abundances). } \end{center}
\end{figure*}

While the metallicity distribution of the outer OCs has often been defined as a plateau, our sample is able to trace a very shallow gradient (Fig.~\ref{fig:gradient_Both}) of \mbox{-0.027$\pm$0.007\,dex\,kpc$^{-1}$}, with a metallicity slowly decreasing over the whole range 12--20\,kpc. However, this apparent trend is mainly driven by one datapoint, as Be~29 is both the most distant and the most metal poor object in the sample, and the gradient could be virtually flat beyond R$_{\mathrm{GC}}\simeq$15\,kpc.

The orbit reconstruction of Sect.~\ref{sec:orbits} identified four clusters (Be~29, Sau~1, Be~20, and Be~75) for which the present-day Galactocentric radius might be significantly different from the birth radius. Be~29 and Sau~1 are the most distant clusters in the sample, and have a large leverage on the observed slope of the gradient. Discarding them leads to a steeper slope of \mbox{-0.063$\pm$0.008\,dex\,kpc$^{-1}$}. On the other hand, discarding the outlier Be~20 only has little influence and yields a value of \mbox{-0.027$\pm$0.007\,dex\,kpc$^{-1}$}. Discarding all four clusters with peculiar orbits and keeping only six datapoints, we find a slope of -0.038$\pm$0.048\,dex\,kpc$^{-1}$.

The slopes we obtain when keeping the outermost two clusters are slightly steeper than values found in literature for the outer disk, for instance -0.02\,dex\,kpc$^{-1}$ outside 13\,kpc \citep{Yong12} or -0.02\,dex\,kpc$^{-1}$ outside 10\,kpc \citep{Frinchaboy13}.
Our values are, however, significantly shallower than the slopes observed in the inner disk, for instance: -0.07\,dex\,kpc$^{-1}$ within 12\,kpc \citep{Andreuzzi11}, -0.06\,dex\,kpc$^{-1}$ within 13\,kpc \citep{Yong12}, \mbox{-0.20$\pm$0.08\,dex\,kpc$^{-1}$} within 10\,kpc \citep{Frinchaboy13}, or -0.08\,dex\,kpc$^{-1}$ within 8\,kpc (Jacobson et al., submitted).

The clusters studied in this paper are intermediate (0.8\,Gyr) to old (5.8\,Gyr) objects. It has been suggested that old clusters may trace a steeper gradient than young clusters \citep[e.g.][]{Friel02,Andreuzzi11,Frinchaboy13}, although the study of \citet{Yong12} reports a larger dispersion but no steeper gradient for old OCs. The chemodynamical model of \citet{Minchev13} suggests that the [Fe/H] gradient was indeed steeper in the past (see in particular their Fig.~5), but because of radial migration the gradient traced by the present-day position of those old tracers should in fact appear shallower than the gradient traced by younger tracers \citep[the same behaviour is reported in][]{Roskar08}. In the model of \citet{Minchev13}, the difference between initial and present-day gradient is negligible ($\sim0.003$\,dex\,kpc$^{-1}$) for objects younger than 2\,Gyr (in this study, Rup~4, Rup~7, Be~73, and possibly Tom~2), but may reach up to a $\sim0.04$\,dex\,kpc$^{-1}$ difference in slope for clusters in the 4--6\,Gyr range (Be~20, Be~29, and Sau~1).
Splitting the sample into two groups, the clusters older than 3\,Gyr (Be~20, Be~22, Be~29, Be~66, Be~75, and Sau~1) trace a gradient of -0.023\,dex\,kpc$^{-1}$, and the younger group (Rup~4, Rup~7, Tom~2, and Be~73) traces a very similar gradient of -0.027\,dex\,kpc$^{-1}$. This sample of ten open clusters is therefore not sufficient to draw conclusions about the time evolution of the metallicity gradient in the outer disk.

The abundance gradient for $\alpha$-elements (average of Mg, Si, Ca, and Ti) appears to be flat, with a slope of +0.006$\pm$0.007\,dex\,kpc$^{-1}$. 
%

The uncertainty on the $\alpha$ abundance for Tom~2 is rather large, as a result of the observations for this cluster having the lowest S/N ratio in the whole sample (see Table~\ref{tab:targets}). Two clusters around R$_{\mathrm{GC}}\sim15$\,kpc appear to be slightly $\alpha$-rich, although with a rather large uncertainty on their $\alpha$ abundances: Be~75 and Be~20. Both of these happen to be among those clusters with irregular orbits (as determined in Sect.~\ref{sec:orbits}). Their position in the [Fe/H] -- [$\alpha$/Fe] plane \citep[see e.g.][]{Bensby03,Haywood15,Kordopatis15} would be at the limit between thin and thick disk, which means that these objects are either genuine old thin disk clusters on perturbed orbits, or members of the younger, metal-rich tail of the thick disk.

The outermost two clusters, Sau~1 and Be~29, also appear slightly $\alpha$-rich, as already reported by \citet{Carraro04} and \citet{Yong05}. Apart from the possibility that these two old clusters are representants of the metal-rich tail of the thick disk, \citet{Yong05} suggests that $\alpha$-enhanced populations might be found in the outskirts of the disk if minor mergers trigger bursts of star formation. 

\begin{figure}[ht]
\begin{center} \resizebox{\hsize}{!}{\includegraphics[scale=0.8]{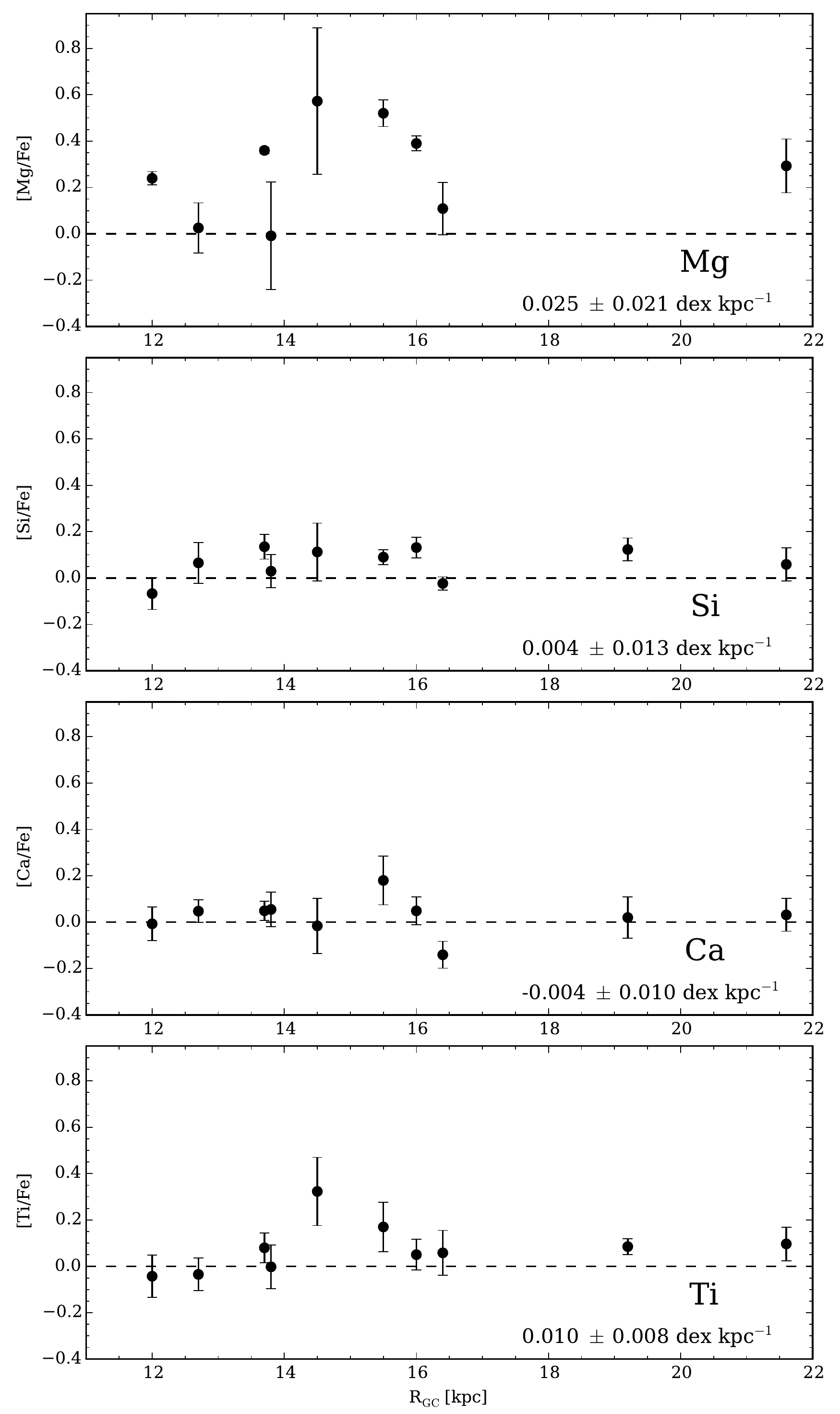}} \caption{\label{fig:4gradients} Radial abundance gradients for Mg, Si, Ca, and Ti, along with the slopes obtained from a linear fit.} \end{center}
\end{figure}

The gradient traced by the individual $\alpha$ elements is shown in Fig.~\ref{fig:4gradients}. A linear fit shows that all distributions are compatible with a flat or slightly positive gradient. The abundances of Mg are based on few (typically two, sometimes three) measured lines, but are all observed to be super-solar. This result corresponds to the observations of \citet{Sestito08} in their study of old outer disk clusters. We remark that if we discard Mg in the computation of $\alpha$-abundances, Be~75 remains the only cluster with a significant $\alpha$-enhancement ([$\alpha$/Fe]=0.15$\pm$0.04). 

\begin{figure*}[ht]
\begin{center} \resizebox{\hsize}{!}{\includegraphics[scale=0.8]{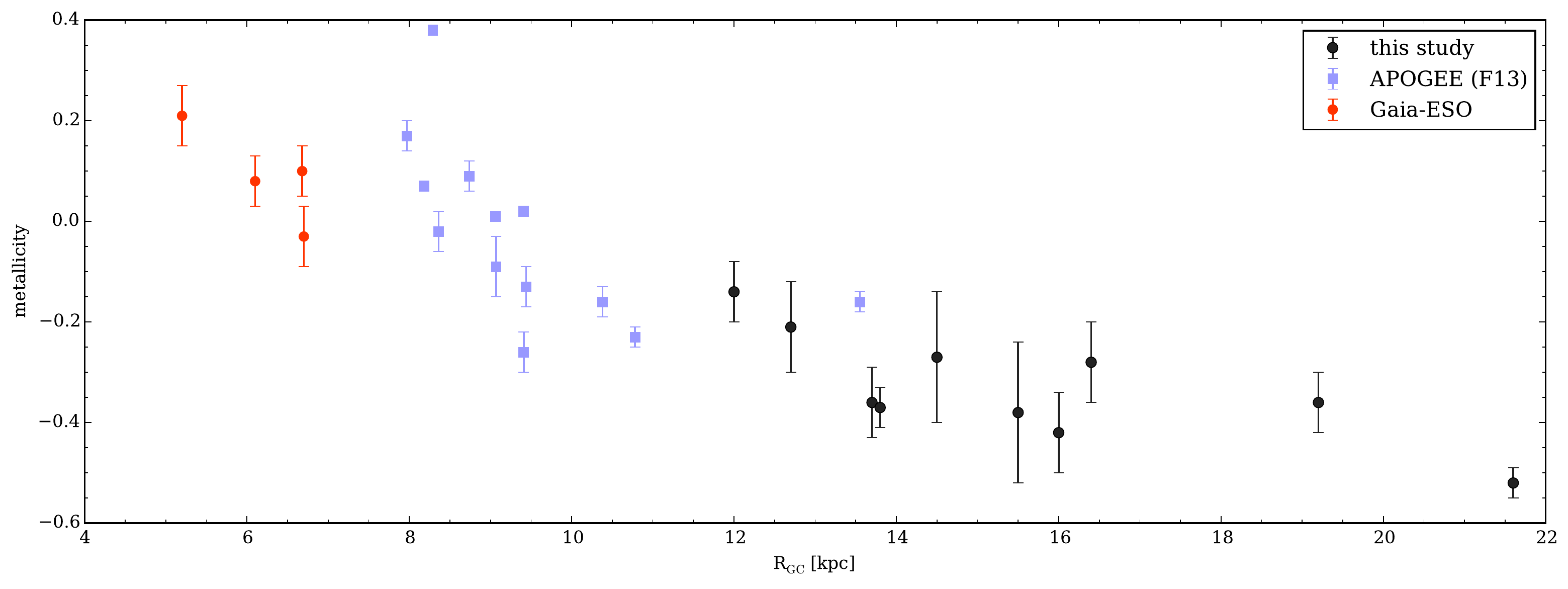}} \caption{\label{fig:metallicity_gradient_withlit} Radial metallicity gradient traced by the first four OCs studied by the Gaia-ESO Survey, the sample of \citet{Frinchaboy13} (F13), and this study. In the sample of F13, we only show clusters with abundances obtained from more than one star.  } \end{center}
\end{figure*}

The sample of OCs under analysis in this study only covers the outer part of the disk and is not sufficient to infer the whole metallicity gradient in our Galaxy. Figure~\ref{fig:metallicity_gradient_withlit} shows, in addition to the results of this paper, the sample of \citet{Frinchaboy13}, \textbf{from which we considered only the clusters with abundances derived from more than one star}, and the abundances of the first four inner-disk clusters studied by GES: Trumpler~20 \citep{Donati14}, NGC~4815 \citep{Friel14}, NGC~6705 \citep{CantatGaudin14m11}, and Berkeley~81 \citep{Magrini15}.
The total gradient traced when piecing together those three samples suggests a constantly declining gradient, with a transition zone rather than an obvious break between an inner-disk and an outer-disk regime. This picture is consistent with the recent results that \citet{Netopil16} obtained from a homogenised sample of spectroscopic and photometric metallicities.

\section{Summary and discussion}
Our chemical abundance determinations show a shallow negative [Fe/H] gradient of slope -0.027$\pm$0.007 dex.kpc$^{-1}$ in the Galactocentric distance range 12 to 20\,kpc. This result seems to indicate that the metallicity does not reach a plateau value, but instead decreases continuously (at a much slower rate than in the inner disk) towards the outside of the disk. We however stress that our sample only contains two clusters beyond 17\,kpc, and that the apparent decrease is driven by a single point (Be~29) which has a strong leverage given its position as the most distant OC from the Galactic centre. If interpreted as a plateau, the flattening around a metallicity of -0.4 is compatible with the results found in other studies \citep[][]{Andreuzzi11,Yong12}.  
Our sample paints a tight relation between Galactocentric radius and metallicity (Fig.~\ref{fig:metallicity_gradient_withlit}). Previous studies, such as the works of \citet{Sestito08} or \citet{Yong12}, reveal a significant spread in metallicity for clusters in the R$_{\mathrm{GC}}$ range 10--13\,kpc. The sample presented in this study only contains two clusters in that distance range, and the sample of \citet{Frinchaboy13} contains another two in that range, which prevents us from studying a possible spread of the R$_{\mathrm{GC}}$--[Fe/H] relation in that region.

A shallower gradient in the outer disk than in the inner disk is predicted by inside-out formation models such as \citet{Magrini09} or \citet{Romano10}, in which the gas infall rate and star-formation rate are higher in the inner disk, leading to a faster chemical enrichment than in the outskirts. It is also predicted by the two-infall evolution model of \citet{Chiappini01}, which considers the bulge and halo formed on a short timescale, while the disk grew inside-out. In the chemodynamical model of \citet{Minchev13}, the disk also grows as a result of gas accretion and gas-rich mergers. A metallicity gradient naturally arises from their simulation, and the present-day slope of this gradient presents a plateau since the stars are scattered by radial mixing. Younger population are expected to have been less affected by radial migration, and their gradient to have been less washed out by radial mixing than older populations. In their model, young tracers are expected to show a steeper gradient. In this study we find no difference between the gradient traced by the older six clusters ($>3$\,Gyr) and by the younger four clusters ($<3$\,Gyr). This result is similar to what is found by \citet{Yong12}, although our study makes use of fewer tracers. 

The abundances of [Mg/Fe], [Si/Fe], [Ca/Fe], and [Ti/Fe] (and globally the value of [$\alpha$/Fe]) appear to be constant across the outer disk, or slightly increasing towards the outskirts, with no significant positive gradient (+0.006$\pm$0.007\,dex\,kpc$^{-1}$). In the simulation of \citet{Minchev13} a shallow positive gradient (no steeper than +0.02\,dex\,kpc$^{-1}$) is expected beyond 12\,kpc, which matches our observations.

We have identified four clusters that appear to follow peculiar orbits (Be~20, Be~29, Be~75, and Sau~1). All four are older than 4\,Gyr, and the most distant clusters from the Galactic plane in our sample (all have $\lvert$z$\rvert$>1.6\,kpc). Their [Fe/H] and [$\alpha$/Fe] mean abundances place them at the limit between the $\alpha$-rich thin disk and $\alpha$-poor thick disk, and their old age and irregular orbits means they could potentially be associated with the thick disk rather than being traditional thin disk clusters. \citet{MartinezMedina16} found that even clusters originating in the thin disk can be sent up to high Galactic altitudes, because of the sharp changes in potential caused by the presence of spiral arms. Their study however focused on the inner 12\,kpc of the disk and the efficiency of this mechanism in the outer regions of the disk is difficult to estimate. 

In the case of Be~29 and Sau~1, which are the two outermost known open clusters in the Milky Way, their peculiar orbits may also be due to perturbations introduced by minor mergers or extragalactic material that is accreted on the outer disk in the past 5\,Gyr (the age of the clusters placing an upper limit on the time of those events). Such a scenario would also explain the slight $\alpha$-enhancement of Sau~1 and Be~29, as mergers are expected to trigger bursts of star formation and lead to higher $\alpha$-abundances \citep{Yong05,Yong06}. Even in the absence of mergers, interactions with satellite galaxies may be responsible for disturbances in the outer disk. The Large Magellanic Cloud is considered one of the main culprits in warping the Galactic disk through tidal interactions \citep{GarciaRuiz02lmc,Weinberg06,Kim14}. The disk of the Milky Way bends southwards in the direction of the anticentre \citep{Levine06warp}, which can partially explain the negative latitudes of Be~20 and Be~75 but makes the high latitudes of Be~29 and Sau~1 even more puzzling.

An alternative and exciting explanation for the observed trajectories of those objects is an extragalactic origin. Those clusters are all located in the direction of the Galactic Anticentre Stellar Structure (GASS; also known as Monoceros stream), first identified by \citet{Newberg02} and confirmed by \citet{Ibata03} as an overdensity of blue stars in the direction of the Galactic anticentre at an estimated distance of 11 -- 16\,kpc from the Sun, using SDSS data. The nature of this structure is still a topic of debate. \citet{Momany06} argue that the Monoceros overdensity can be explained in terms of a both warped and flared disk. 
\citet{Ivezic08} determined that the kinematics of the GASS are very close to the behaviour of the Galactic disk, giving weight to the surmise that the GASS stars are all Milky Way stars. \citet{Juric08} reach an opposite conclusion and find that the Monoceros structure has the shape of a stream, and would be a remnant of a dwarf satellite galaxy cannibalised by the Milky Way. A review on this decade-long debate was published by \citet{LopezCorredoira12}. Multiple studies have attempted to characterise the chemistry of the GASS, typically obtaining rather low metallicities: [Fe/H]=-1.6 \citep{Yanny03}, -1 \citep{Meisner12,Conn12}, -0.95 \citep{Martin04}, -0.8 \citep{Li12gass}, -0.4 \citep{Crane03}. 

Making use of radial velocity measurements, \citet{Frinchaboy04} and \citet{Frinchaboy06} investigated the possibility that some outer disk clusters are the product of the accretion of a dwarf satellite galaxy. They conclude that Be~29 and Sau~1 have RVs compatible with the GASS. Be~20 is located right in the middle of the region of the hypothetical Monoceros structure. Despite the fact that these three objects are the most metal poor in our sample, we still find that their [Fe/H] abundances of -0.52, -0.36 and -0.42 (for Be~29, Sau~1 and Be~20, respectively) are more compatible with them being Galactic clusters than with the estimated abundances of Monoceros.

 A study by \citet{Carraro09} investigated the possibility that Be~29 and Sau~1 are associated with the Sagittarius dwarf spheroidal galaxy \citep[or Sgr dSph,][]{Ibata94}, which is currently being accreted on the Galactic disk, rather than with Monoceros. They showed that their position and kinematics are compatible with an origin within Sgr dSph, and their [Fe/H] and [$\alpha$/Fe] abundances are marginally compatible with being in the metal-rich tail of the Sgr dSph population, although the Ca abundances of those objects do not match the subsolar [Ca/Fe] reported by \citet{Monaco07} and \citet{Sbordone07} for Sgr dSph. The two known dwarf galaxies with metallicities above -1 (Sgr dSph and Fornax) in fact exhibit subsolar $\alpha$ abundances of about $\sim$-0.2\,dex \citep[see e.g.][and references therein]{Tolstoy09}, making them unlikely candidates for the origin of our outer-disk open clusters.

Our study does not reveal any significant chemical anomalies among the clusters of the outer disk, and therefore there is no strong evidence that any of them may be linked to peculiar structures or originate from outside the Galaxy. We do however show evidence that some clusters are potentially associated with the Galactic thick disk (Be~20, Be~75), or are indicators of merger events having affected the outer disk in the past 5\,Gyr (Be~29, Sau~1).
We expect significant progress to be made in the identification of structures and characterisation of their kinematics in the years to come thanks to the contribution of the Gaia mission, which will provide high-quality proper motions for stars down to $V$=20, allowing us to probe the properties of the outer disk and find traces of past, and ongoing, accretion events.

\section*{Acknowledgements}
Based on data obtained from the ESO Science Archive Facility under request numbers 160723, 165286, 165299, 165302, 165304, 165305, 165306, 165307, 165466, and 165467.
Some of the data presented herein were obtained at the W.M. Keck Observatory, which is operated as a scientific partnership among the California Institute of Technology, the University of California and the National Aeronautics and Space Administration. The Observatory was made possible by the generous financial support of the W.M. Keck Foundation. The authors wish to recognise and acknowledge the very significant cultural role and reverence that the summit of Mauna Kea has always had within the indigenous Hawaiian community.  We are most fortunate to have the opportunity to conduct observations from this mountain.

We acknowledge the support from INAF and Ministero dell'Istruzione, dell'Universit{\'a} e della Ricerca (MIUR) in the form of the grant ``Premiale VLT 2012" and ``The Chemical and Dynamical Evolution of the Milky Way and the Local Group Galaxies" (prot. 2010LY5N2T).  This research has made use of WEBDA, SIMBAD database, operated at CDS, Strasbourg, France, and NASA's Astrophysical Data System.


\bibliographystyle{astron_tristan} 
\linespread{1.5}		
\bibliography{biblio}

\end{document}